\documentclass{article}

\usepackage{jheppub}

\usepackage[utf8]{inputenc}
\usepackage{verbatim}
\usepackage{natbib}
\usepackage{graphicx}
\usepackage{amsmath,amssymb}
\usepackage{dsfont}
\usepackage{cancel}
\usepackage{float}
\usepackage{braket}
\usepackage{siunitx}
\usepackage{xspace}
\usepackage[capitalize]{cleveref}
\usepackage{comment}
\usepackage[usenames, dvipsnames]{color}
\usepackage{fancyvrb}

\binoppenalty=\maxdimen
\relpenalty=\maxdimen

\makeatletter
\gdef\@fpheader{\strut}
\makeatother

\crefname{section}{Sec.}{Secs.}

\newcommand*{\Mpci}{\ensuremath{\ \mathrm{Mpc}^{-1}}}
\newcommand*{\mPl}{\ensuremath{m_\mathrm{Pl}}\xspace}
\newcommand*{\vect}[1]{\ensuremath{\mathbf{#1}}\xspace}

\newcommand*{\Neff}{\ensuremath{N_{\mathrm{eff}}\xspace}}
\newcommand*{\dtc}{\ensuremath{\dot\tau_C}}
\newcommand*{\aeq}{\ensuremath{a_{\rm{eq}}}}
\newcommand*{\Pow}{\ensuremath{\mathcal{P}}}

\newcommand{\thetan}[1]{\ensuremath{\theta_{n,#1}}}

\newcommand{\dthetan}[1]{\ensuremath{\dot\theta_{n,#1}}}

\preprint{MITP-22-077}

\title{One $\mu$ to rule them all: CMB spectral distortions can probe domain walls, cosmic strings and low scale phase transitions}

\author[a]{Nicklas~Ramberg,}
\author[a]{Wolfram~Ratzinger}
\author[a]{and Pedro~Schwaller}

\affiliation[a]{PRISMA$^+$  Cluster of Excellence and Mainz Institute for Theoretical Physics, Johannes  Gutenberg-Universit\"at  Mainz, 55099 Mainz, Germany}

\emailAdd{nramberg@uni-mainz.de}
\emailAdd{w.ratzinger@uni-mainz.de}
\emailAdd{pedro.schwaller@uni-mainz.de}

\abstract{We present a new probe of purely gravitationally coupled sectors with large anisotropies. These anisotropies are damped via gravitational interactions with the baryon-photon fluid, which is heated up in the process. The injected heat causes measurable distortions of the cosmic microwave background spectrum. We give analytic estimates for the size of the distortions and outline how to calculate them from first principles. These methods are applied to anisotropies in the form of a domain wall/cosmic string network or caused by a first order phase transition or scalar field dynamics. We find that this method can potentially probe large regions of previously unconstrained parameter space and is very much complementary to up-coming searches of gravitational waves caused by such dark sectors.

}

\begin{document}

\maketitle\flushbottom

\section{Introduction}
%

Dark sectors that only interact with our visible Universe gravitationally are well motivated - they often arise in string theory compactifications, or more generally appear in models with hidden sectors that address the dark matter or other puzzles of the standard model (SM). 
Gravitational waves have emerged as a prominent new way to probe such dark sectors. This is possible if the dark sector features large anisotropies , which then compensate the intrinsic weakness of the gravitational interactions. Sizeable anisotropies can be present in the dark sector in the form of topological strings or domain walls~\cite{Kibble:1976sj,Vilenkin:1982ks,Vilenkin:1984ib}, 
or can be produced for example in first order phase transitions~\cite{Hogan:1983ixn,Witten:1984rs,Kuzmin:1985mm} or by scalar field dynamics~\cite{Machado:2018nqk,Fonseca:2019ypl,Chatrchyan:2020pzh}.

Here we present a new way to detect the presence of such large anisotropies in a dark sector via their effect on the spectrum of the cosmic microwave background (CMB). The dark sector anisotropies are damped via gravitational interactions with the baryon-photon fluid, which is heated up in the process. If this happens shortly before the CMB is emitted, the photons do not have enough time to re-thermalise, and this process leads to a deviation of the CMB from a perfect black body spectrum~\cite{Zeldovich:1969ff,Chluba:2011hw,Kogut:2019vqh}.

These spectral distortions of the CMB are tightly constrained by existing observations, and can therefore be used to probe dark sector models that feature large anisotropies around the time of CMB emission. The constraints we obtain below are based on measurements dating back to the nineties, but already provide competitive bounds on the parameter space of some scenarios. With current technology, they could be improved by about four orders of magnitude, and we show that such measurements would dig deep into previously unconstrained parameter regions of dark sectors.

Dark sector anisotropies induce both acoustic waves in the baryon-photon fluid as well as gravitational waves (GWs). The latter also lead to spectral distortions, as was shown in Ref.~\cite{Chluba:2019nxa}. We find however that there is a large class of dark sector models for which the contribution from acoustic waves dominates, and thus our mechanism often leads to significantly stronger constraints. 
In the case of non-decaying cosmic string and scaling seed networks the spectral distortions caused by acoustic waves were already discussed in \cite{Tashiro:2012pp,Amin:2014ada}. We discuss this phenomenon in general here and provide estimates applicable to a plethora of dark sectors. Besides the aforementioned cosmic strings, this includes models with domain walls, very late first order phase transitions, or scalar fields undergoing parametric resonance.

In \cref{sec:CMB-distortions} we provide an overview of $\mu$-distortions and qualitatively discuss the conditions under which dark sectors can source observable distortions. 
 From there, the reader only interested in the reach of this new probe is advised to jump to \cref{sec:GW_sources}, where we apply our techniques to various dark sectors that are well-known GW sources. \cref{sec:source_mu,sec:sound_energy,sec:lambdaphi4} instead provide more technical details: 
In \cref{sec:source_mu} we show that the induction of sound waves in the photon fluid through the dark sector can be decoupled from the subsequent damping, and derive an expression for the resulting $\mu$-distortions. 
Analytic results for the acoustic energy caused by different types of dark sectors are obtained in~\cref{sec:sound_energy}. Then in \cref{sec:lambdaphi4} we put our analytical estimates to the test by comparing them to numerical results of a dark sector toy model.

\section{CMB spectral distortions}
\label{sec:CMB-distortions}

It is well known that the CMB spectrum is to a good approximation a black body spectrum. Any deviation of the spectrum from this shape, so called spectral distortions, however, encode valuable information about physics in the early universe. In principle any non-thermal injection or removal of energy from the photons causes such a distortion. Whether a distortion is observable depends, aside from the size of the distortion, on whether efficient processes to thermalize the spectrum again are present.

At high red-shift, and correspondingly large temperatures, processes changing the photons momentum, like Compton scattering, as well as photon number changing processes, like double Compton scattering, are present. In this regime any distortions of the CMB are therefore quickly erased. 
This changes for redshifts $z\lesssim 2\times 10^6$ when photon number changing processes become ineffective. From this point onwards the photon number is a conserved quantity and one has to introduce a chemical potential $\mu$ to capture the equilibrium distribution. It becomes non-zero if energy is injected into the plasma at this point. Below red-shifts of $z\lesssim 5\times 10^4$ Compton scattering also becomes inefficient at redistributing the momentum among the photons, such that any distortion sourced at later times is directly imprinted onto the CMB spectrum.

The source of energy injection we are interested in here is the damping of sound waves in the baryon-photon fluid. A sound wave in the plasma with momentum $k$ is rapidly damped once its wavelength $\lambda = 2 \pi/k$ falls below the diffusion scale, which is the distance a photon covers by random walking between scattering events. For modes in the range $8 \times 10^3\ \text{Mpc}^{-1}\lesssim k\lesssim 2\times 10^3\ \text{Mpc}^{-1}$ this happens for redshifts $5\times 10^4 \lesssim z\lesssim 2\times 10^6$ leading to a $\mu$-distortion.
\footnote{For larger wavelength modes where the damping occurs for $z\lesssim 5\times 10^4$ the photon momenta are not redistributed resulting in a $y$-distortion. This is also produced through the Sunyaev-Zeldovich effect \cite{Zeldovich:1969ff,Chluba:2011hw} limiting the sensitivity to new physics. Our results for $\mu$-distortions can be generalized by using the respective window function (see \cref{sec:source_mu} and \cite{Chluba:2013dna}).}

In the inflationary paradigm the primordial fluctuations, measured at the largest scales as CMB fluctuations and in structure formation, are predicted to be approximately flat and therefore extend to the small scales sourcing $\mu$-distortions. In this paper we however investigate dark sectors with turbulent dynamics which, through their gravitational coupling to the photon fluid, lead to additional fluctuations at those small scales and therefore source additional $\mu$-distortions, while leaving the scales relevant for CMB fluctuations and structure formation untouched.

    \begin{figure}
        \centering
        \includegraphics[width=0.8 \textwidth]{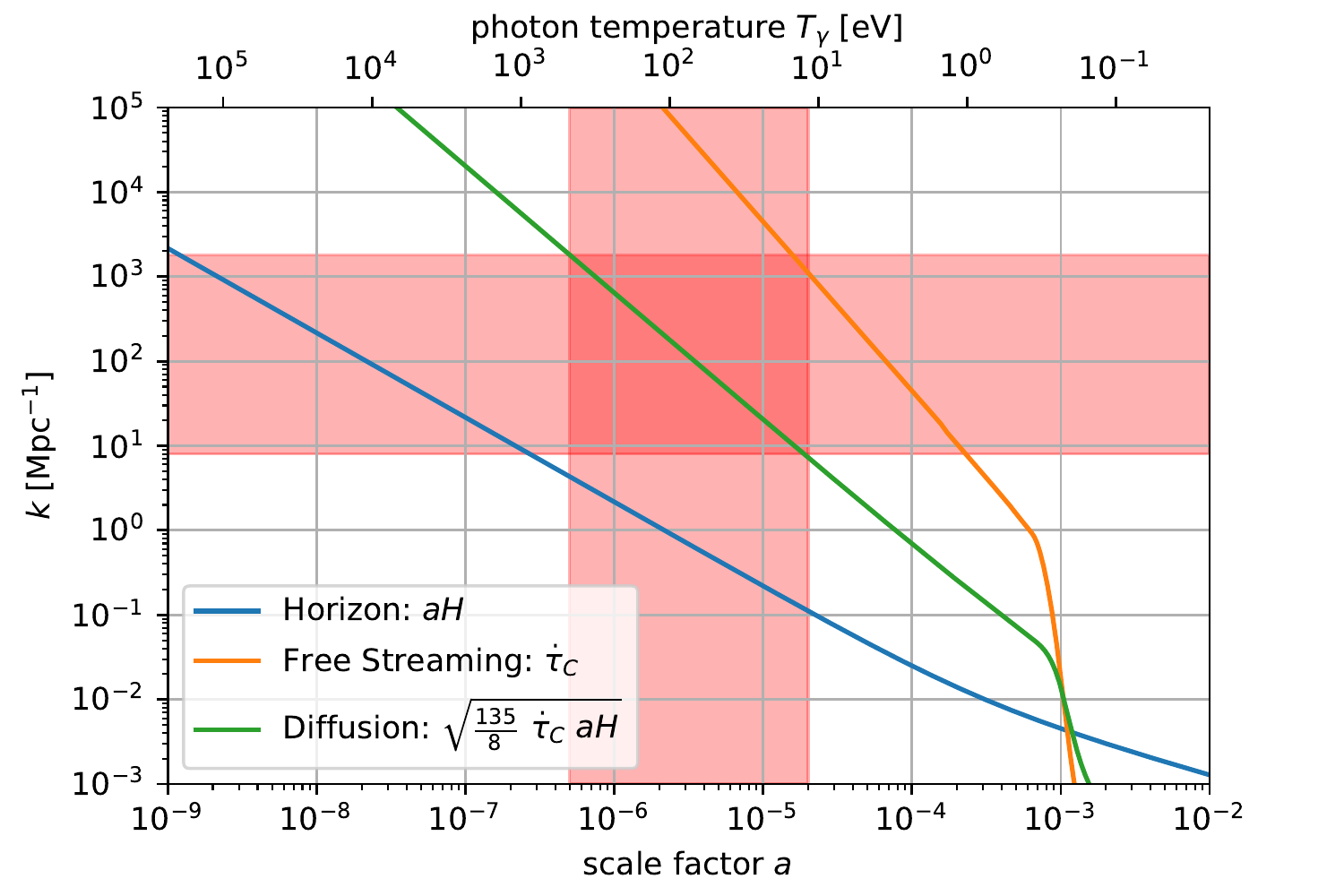}
        \caption{The evolution of the horizon scale (blue), the scale a photon free streams between consecutive scattering events and diffusion scale that is approximately the distance traveled by a photon in a random walk as a result of all the scattering events in one Hubble time. Once a mode passes the diffusion scale the energy stored in the acoustic oscillation  is damped. If this happens during red-shifts marked in red on the x-axis a $\mu$-distortion is sourced, singling out the modes marked in red on the y-axis as the dominant messengers of new physics. These modes enter the Horizon as early as $a\approx10^{-9}$ allowing us to search for new physics back to when the photon temperature was $\approx 1-0.1\ \text{MeV}$. }
        \label{fig:scales}
    \end{figure}

The details of the gravitational coupling between the different sectors are discussed below. We find that the coupling is strongest around the time a mode crosses the horizon (blue in \cref{fig:scales}). Past this point the energy of the acoustic wave in the baryon-photon fluid is approximately constant until the mode crosses the diffusion scale (green in \cref{fig:scales}) at which point the acoustic energy is turned into bulk energy. We have highlighted on the $x$-axis of \cref{fig:scales} the times at which this injection of energy leads to a $\mu$-distortion. The modes that cross the diffusion scales during these times, highlighted on the $y$-axis, are therefore the dominant messengers of new physics. We can read off that these modes enter the horizon as early as $a\approx 10^{-9}$ corresponding to photon temperatures of $\approx 0.1-1$~MeV. Indeed we find in \cref{sec:GW_sources} that the sensitivity of spectral distortions is diminished for scenarios in which the anisotropies appear at higher temperatures. It is conceivable that energy is injected into high $k$ modes that are already past the damping scale or even free streaming scale and converted into bulk energy of the plasma in a fraction of a Hubble time. However we find that this process is highly inefficient due to the large separation between the damping and free streaming scale and the horizon at the times of interest.

Since the interaction between the different sectors is purely gravitationally, we find, perhaps unsurprisingly, that the induced spectral distortions are proportional to the energy comprised by the dark sector $\Omega_d$. Furthermore the amplitude $\delta_d$ and characteristic scale $k_*$ of the fluctuations determine the magnitude of the distortions. Larger fluctuations clearly lead to bigger distortions. On the other hand a miss-match between the characteristic scale $k_*$, where the fluctuations are largest, and the Horizon, where the gravitational coupling is strongest, leads to a reduction of the distortions. An example for the last factor would be a first order phase transition, where the bubble radius at collision can be much smaller than the Hubble radius.

\section{Source of $\mu$-distortions through gravitational interaction}
\label{sec:source_mu}
The generic setup we have in mind is a dark sector that only comprises a subdominant amount of the total energy $\Omega_{d} \ll 1$ but develops large anisotropies at some point $\delta\rho_{d}/\overline\rho_{d}=\delta_d\approx 1$.
Since $\Omega_{d} \ll 1$, metric and density fluctuations in other sectors present in the universe remain small, which allows us to treat them perturbatively, linearizing their dynamics. It is then beneficial to work in Fourier space. We use the following definitions for the Fourier transform and the dimensionless power spectrum $\mathcal{P}$:
\begin{align}
    \phi(\vect{k})=\int d^3x\ \phi(\vect{x})\exp(-i\vect{kx})\,;\qquad \langle \phi(\vect{k})\phi^*(\vect{k}') \rangle=\frac{2\pi^2}{k^3}\mathcal{P}_\phi(k)(2\pi)^3\delta^{(3)}(\vect{k}-\vect{k}')\,.
\end{align}
We furthermore use the conformal Newtonian gauge for the scalar metric perturbations
\begin{align}
    ds^2=a^2(\tau)\left[(1+2\Psi(\vect{x},\tau))d\tau^2-(1+2\Phi(\vect{x},\tau))d\vect{x}^2\right],
\end{align}
where the equations of motion for the potentials $\Phi,\Psi$ are given by
\begin{align}
    3aH\left(\dot{\Phi}-aH\Psi \right) + k^{2}\Phi &= \frac{3a^{2}H^2}{2}\left(\Omega_{\gamma}\delta_{\gamma} + \Omega_{n}\delta_{n} + \Omega_{d}\delta_{d} \right)\label{eq:phi}\,,\\
        \Phi + \Psi &= -\frac{6a^{2}H^2}{k^{2}}\left(\Omega_{\gamma}\sigma_{\gamma} + \Omega_{n}\sigma_{n} +\frac{3}{4}(1+w_d) \Omega_{d}\sigma_{d}\right).\label{eq:psi}
\end{align}
Here overdots denote derivatives w.r.t conformal time $\tau$, and $\delta$ and $\sigma$ denote the energy fluctuation and shear in the respective sectors. For times well before matter-radiation equality, $a\ll \aeq$, the three relevant sectors are the baryon-photon fluid $\gamma$, the neutrinos $n$ and the dark sector $d$. The shear is defined as the longitudinal traceless part of the energy-momentum tensor $\sigma=-(\vect{\hat k}_i\vect{\hat k}_j-\frac{1}{3}\delta_{ij})T^i_j/(\overline\rho+\overline p)$. Finally, $w_d$ is the equation of state parameter of the dark sector.

It is clearly visible that the scalar metric perturbations induced by $\delta_d$ are suppressed by $\Omega_d$, thus justifying the linearised treatment. More generally the linearized treatment holds as long as either $\Omega_d$, $\delta_d$ or $H^2/k^2$ is small and none of them are larger than 1.  Furthermore since all other sectors only couple to the dark sector via gravity, also their induced perturbations are suppressed. This also allows one to neglect the back-reaction effects of gravity onto the dark sector and one can therefore study its dynamics using an unperturbed metric. In the following, we will assume that all fluctuations in the baryon-photon fluid and the neutrinos as well as the potentials are initially zero. 
The effects of other uncorrelated fluctuations like e.g. inflationary ones can be studied independently, as usual in linear perturbation theory.

For modes deep inside the horizon, $k\gg aH$, one can solve for the gravitational potentials directly by neglecting the first term on the right side of \cref{eq:phi}.
One finds that the gravitational potentials decay as $\Phi,\Psi\propto a^2H^2/k^2$ if the fluctuations don't keep growing after their generation, which is a reasonable assumption during radiation domination. Therefore the gravitational coupling between the sectors quickly becomes negligible after horizon entry. For our specific case, it suggests that the amplitude of fluctuations in the baryon-photon fluid is set within about one Hubble time after horizon entry or after the dark sector fluctuations have been created, whichever happens later for a given mode. We can also anticipate the strength of the gravitational interaction being suppressed for modes that are deep inside the horizon when the dark sector develops its fluctuations. The details of this suppression are discussed in the next section.

For the times well before recombination, the baryon-photon fluid is well described by the tight-coupling approximation (TCA, e.g. \cite{Hu:1995en,Ma:1995ey}) and the energy density in baryons can be neglected, leading to
\begin{align}
    \dot\delta_\gamma + \frac{4}{3} k v_\gamma&=-4\dot\Phi\label{eq:delta_gamma}\\
    \dot v_\gamma- k\left(\frac{1}{4} \delta_\gamma-\sigma_\gamma\right)&=k\Psi\label{eq:v_gamma}\\
    \sigma_\gamma&=\frac{16}{45}\frac{k}{\dtc}v_\gamma,
\end{align}
where $v_\gamma=i\vect{\hat k}_iT_{\gamma,0}^{j}/(\overline\rho_\gamma+\overline p_\gamma)$ is the longitudinal part of the fluid velocity relative to the cosmological rest frame. The TCA takes advantage of the fact that all moments of the photon distribution past the velocity are suppressed by the high Compton scattering rate $\dtc=a n_e \sigma_C \gg k$, where $\sigma_C$ is the Compton cross-section and $n_e$ the free electron density, which can be approximated as $\dtc=a^{-2}\ 4.5\times 10^{-7}\ \text{Mpc}^{-1}$ well before recombination. When solving these equations numerically we also take into account the free streaming neutrinos, see \cref{sec:TCA_nuDyn}.

The equations above can be combined to get a damped harmonic oscillator. We will first do so in the limit that the mode is already deep inside the horizon and neglect the gravitational potentials
\begin{align}
    \ddot\delta_{\gamma}+k^2\left(\frac{16}{45}\frac{1}{\dtc}\dot\delta_{\gamma}+\frac{1}{3}\delta_{\gamma}\right)=0\,.
\end{align}
In the given limit that $aH\ll k \ll \dtc$ the general solution to this problem is approximated as
\begin{align}
    \delta_\gamma=\left[A \sin\left(\frac{k\tau}{\sqrt{3}}\right)+B \cos\left(\frac{k\tau}{\sqrt{3}}\right)\right]\exp\left(-\frac{k^2}{k^2_D(\tau)}\right).
\end{align}
This solution is interpreted as damped acoustic waves traveling in the baryon photon fluid with the relativistic speed of sound of $c_s=1/\sqrt{3}$. The diffusion scale $k_D$ appearing here is determined by the equation $\frac{d}{d\tau}k^{-2}_D=\frac{8}{45}\frac{1}{\dtc}$. During radiation domination it is given as $k_D=\sqrt{\frac{135}{8}\dtc\ aH}$ as long as the free electron density is constant up to dilution by expansion. This effect is also known as Silk damping \cite{Hu:1995kot} and is attributed to photons performing a random walk with typical step length $\Delta x\approx1/\dtc$ while doing $N\approx \frac{\dtc}{aH}$ steps per Hubble time. The diffusion scale is then the distance typically traversed by a photon in a Hubble time, $1/k_D\approx \Delta x\sqrt{N}$. Fluctuations on scales smaller than $1/k_D$ are therefore quickly erased. The energy of the acoustic waves is converted into photon bulk energy in this process.

From the discussion so far two important scales have emerged: the Horizon scale at which the gravitational coupling is strongest and we, therefore, expect the dark sector to efficiently source acoustic waves, and the diffusion scale. As can be seen in \cref{fig:scales}, the modes of interest, marked in red on the y-axis, pass these two scales at scale factors $a$ that are always separated by about two orders of magnitude or more. This means that we can separately discuss the generation of acoustic waves from dark sector anisotropies and the conversion of these acoustic waves into a $\mu$ distortion. 
After they are generated, but before the onset of damping, the amplitudes of the sound waves $A,B$, and therefore also the energy in acoustic waves, remains approximately constant. 

In the remainder of this section we obtain an expression for the $\mu$ distortion generated by the diffusion of the acoustic waves, while the computation of the acoustic energy induced by different sources is postponed to the following sections. 
Relative to the total energy in the relativistic baryon-photon fluid, the acoustic energy is given as
\begin{align}
    \epsilon_{ac}=\frac{\rho_{ac}}{\overline\rho_\gamma}=\frac{1}{V}\int_V d^3x\ \left[\frac{1}{8} \delta_\gamma^2(\vect{x})+\frac{2}{3} v_\gamma^2(\vect{x})\right]=\int d\log k\ \epsilon_{ac}(k)\,,
\end{align}
where we defined the spectral acoustic energy in the last step which is given as
\begin{align}\label{eq:ac_energy}
    \epsilon_{ac}(k)=\frac{1}{8} \mathcal{P}_{\delta_\gamma}(k)+\frac{2}{3} \mathcal{P}_{v_\gamma}(k)=\frac{1}{8}\left[\mathcal{P}_{A}(k)+\mathcal{P}_{B}(k)\right]
\end{align}
in terms of the power spectra for $\delta_\gamma, v_\gamma$ and $A,B$ respectively.

When the acoustic waves get damped by diffusion, this energy becomes part of the photon bulk energy. If this happens between $a_{dc}=5\times 10^{-7}$, when photon number changing processes such as Double Compton scattering becomes inefficient, and $a_{\mu,y}=2\times 10^{-5}$, when Compton scattering stops redistributing the momentum between the photons, then a $\mu$-distortion gets sourced besides an increase in the bulk temperature. The approximation commonly used to determine the $\mu$-parameter is
\begin{align}\label{eq:mu_from_ac_power}
    \mu\approx 1.4\int d\log k \int_{a_{\mu,y}}^\infty d\log a\ \frac{d\epsilon_{ac}(k)}{d\log a}\exp\left(-\left(\frac{a_{dc}}{a}\right)^{5/2}\right),
\end{align}
where $\frac{d\epsilon_{ac}(k)}{d\log a}$ is the acoustic ``power'' transmitted to the bulk energy and it is given as \cite{Hu:1995en}
\begin{align}\label{eq:ac_power}
    \frac{d\epsilon_{ac}(k)}{d\log a}=\frac{15}{4} \frac{\dtc}{aH}\mathcal{P}_{\sigma_\gamma}=\frac{64}{135}\frac{k^2}{\dtc aH}\mathcal{P}_{v_\gamma}=\frac{8}{3}\mathcal{P}_{v_\gamma}\frac{d}{d\log a}\left(\frac{k^2}{k_D^2}\right)\approx2\epsilon_{ac}(k,a)\frac{d}{d\log a}\left(\frac{k^2}{k_D^2}\right),
\end{align}
where we used in the last step that due to the oscillation time scale being much shorter than the damping time scale $1/k\ll1/k_D$ one can approximate $\epsilon_{ac}\approx\frac{4}{3} \mathcal{P}_{v_\gamma}$ in \cref{eq:ac_energy}. In the limit that the acoustic energy takes on a constant value $\epsilon^{\lim}_{ac}(k)$ before damping, we have $\epsilon_{ac}(k,a)= \epsilon^{\lim}_{ac}(k)\exp(-2k^2/k^2_D(a))$ during the period of damping, such that we can write the $\mu$-parameter as
\begin{align}\label{eq:mu_from_W}
    \mu=\int d\log k\ \epsilon^{\lim}_{ac}(k) \mathcal{W}(k),
\end{align}
where we have introduced the window function \cite{Chluba:2013dna}
\begin{align}\label{eq:W_analytic}
    \mathcal{W}(k)&\approx 1.4 \int_{a_{\mu,y}}^\infty d\log a\ \exp\left(-\left(\frac{a_{dc}}{a}\right)^{5/2}\right) \frac{d}{d\log a} \exp\left(-2\frac{k^2}{k^2_D(a)}\right)\\
    &\approx 1.4\left(\exp\left[-\left(\frac{k}{1360\Mpci}\right)^2\left(1+\left(\frac{k}{260\Mpci}\right)^{0.3}+\frac{k}{340\Mpci}\right)^{-1}\right]\right.\\
    &\hspace{1.5cm}\left.-\exp\left[-\left(\frac{k}{32\Mpci}\right)^2\right]\vphantom{\left[\left(\left(\frac{1}{1}\right)^{0.3}\right)^{-1}\right]}\right).\nonumber
\end{align}
This remarkably easy expression allows one to calculate the $\mu$-distortion a dark sector causes, given the spectral acoustic energy before damping.\footnote{In the literature the window function is commonly defined with respect to a primordial spectrum rather than the acoustic energy spectrum and therefore represents a convolution of the dynamics of horizon entry and damping (e.g. \cite{Chluba:2013dna}). Our definition is universally applicable, although one would have to discuss horizon entry separately.}
This value can then be compared to current bounds and the detection threshold of future experiments.

    \begin{figure}
        \centering
        \includegraphics[width= \textwidth]{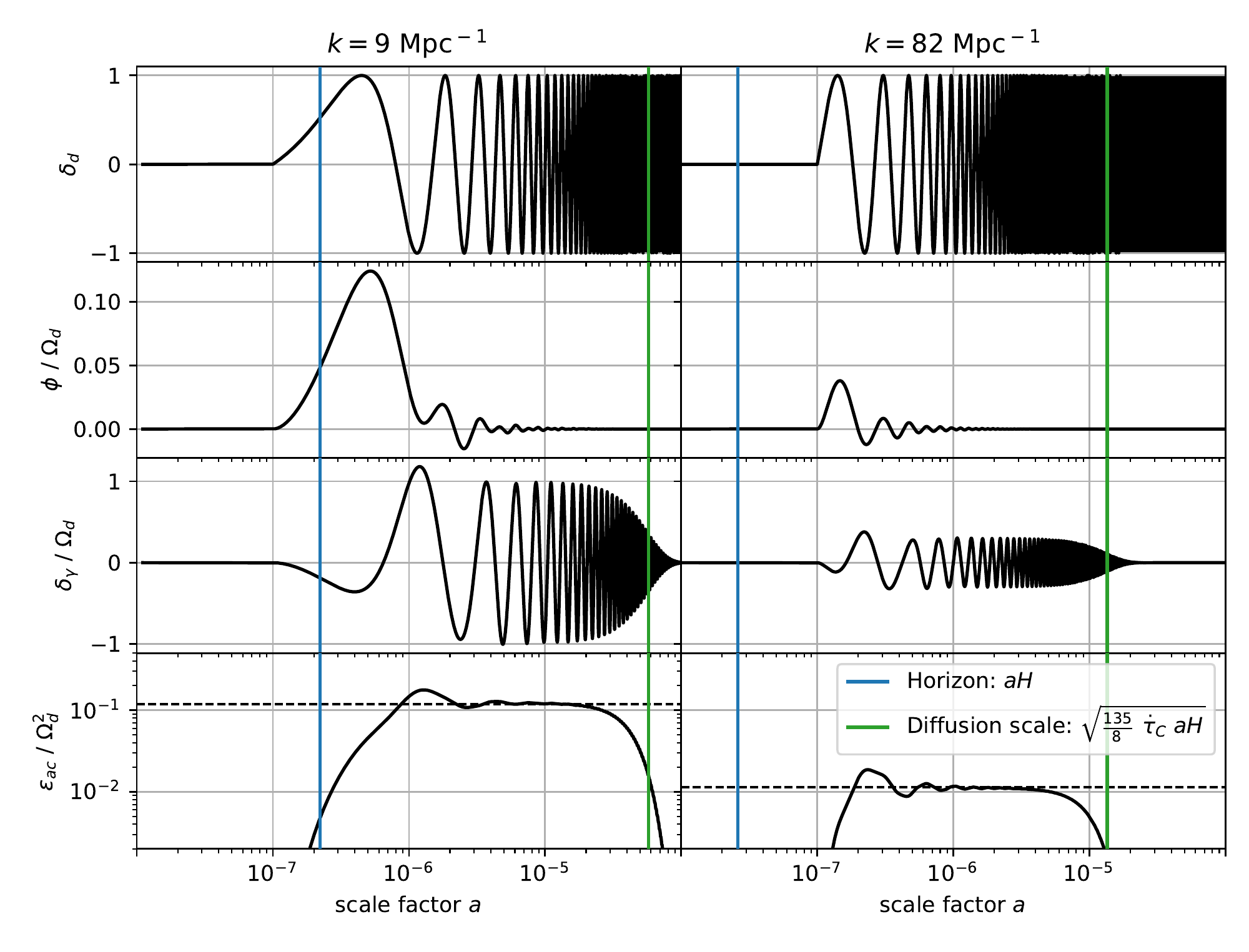}
        \vspace*{-2\baselineskip}
        \caption{Sourcing of acoustic waves through a gravitationally coupled dark sector and subsequent damping by diffusion. The toy dark sector here is radiation like ($\Omega_d=\text{const.}\ll1$) and its density fluctuations are zero until $a_*=10^{-7}$ before evolving as $\delta_d=\sin(k\tau)$ (top row). The resulting gravitational potential (second row) causes acoustic oscillations in the baryon-photon fluid (third row). Since the gravitational potential rapidly decays after a mode has entered the horizon (vertical blue line) the amplitude of the acoustic oscillations quickly levels off resulting in an approximately constant acoustic energy $\epsilon^{\lim}_{ac}(k)$ (bottom row, dashed line). The same effect also leads to the acoustic oscillations being suppressed for the high $k$ mode (right side) that is already inside the horizon when the fluctuations in the dark sector develop. Finally the acoustic oscillations are damped for both modes once they cross the diffusion scale (green line). At this point the acoustic energy is injected into the bulk photon energy, leading to a sizeable $\mu$-distortion for modes where this happens between $5\times 10^{-7}\lesssim a\lesssim 2\times 10^{-5}$.}
        \label{fig:toy_mode_evolution}
    \end{figure}

In \cref{fig:toy_mode_evolution} we show the numerical results of a toy model that neatly summarize this section. The dark sector is assumed to be radiation like such that $\Omega_d=\text{const.}$ while the sound waves are sourced, and the fluctuations are modeled as being zero until $a_*=10^{-7}$ and as $\delta_d=\sin(k\tau)$ afterwards.\footnote{As we will argue in \cref{sec:num_results}, $\dot\delta_d(k\tau)$ only changes on time scales $1/k$ for sub-horizon modes making this an unphysical choice with $\dot\delta_d(k\tau)$ being discontinuous at $a_*$. We only use this ansatz here for demonstration as well as for rough estimates in the following section. } The shear in the dark sector is set to zero. We show the evolution for one mode that is still outside the horizon at $a=a_*$ and one that is already inside. Indeed it can be clearly seen that the acoustic energy is constant between generation and damping, justifying our separation approach.

\section{Analytic Estimate of the Induced Acoustic Energy}
\label{sec:sound_energy}
We now obtain an analytic estimate for the acoustic energy $\epsilon^{\lim}_{ac}$ caused by fluctuations in a dark sector, which together with the results from the previous section allows us to compute the $\mu$ distortions. 
We assume that the fluctuations are generated at a fixed time $a=a_*$. For modes that enter the horizon around or after $a_*$ ($k\lesssim a_*H_*$), we find that the contribution of the photons and neutrinos to the gravitational potentials is of the same order as the one from the dark sector ($\Omega_d\delta_d\approx\Omega_\gamma\delta_\gamma\approx \Omega_n\delta_n$). The coupled system of equations can therefore only be solved numerically. Instead for modes that already are inside the horizon, $k>a_*H_*$, the amplitudes of $\delta_\gamma$ and $\delta_n$ remain suppressed by some power of $a_*H_*/k$, as discussed above. They can therefore be neglected when solving for the gravitational potentials. We therefore restrict our analytic treatment to $k>a_*H_*$.

To make further progress we again combine \cref{eq:delta_gamma} and \cref{eq:v_gamma} but this time keeping the potentials and dropping the diffusion damping, since we now want to solve for times well before the mode crosses the damping scale, to find
\begin{align}
    \ddot\delta_{\gamma}+\frac{1}{3}k^2\delta_{\gamma}&=-4\ddot\Phi-\frac{4}{3}k^2\Psi\,.
\end{align}
To get rid of the second time derivative of $\Phi$, we define $\tilde\delta_\gamma=\delta_{\gamma}+4\Phi$. As the gravitational potential decays, at late times we have $\tilde\delta_\gamma\approx\delta_{\gamma}$. Since we consider a sub-horizon mode, we can continue with only the last term on the left-hand side of \cref{eq:phi}, which allows us to solve for the potentials in terms of $\delta_d$ and $\sigma_d$ directly: 
\begin{align}
    \ddot{\tilde\delta}_{\gamma}+\frac{1}{3}k^2\tilde\delta_{\gamma}&=4a^2H^2\ \Omega_{d}\left[\delta_{d} + \frac{3}{2}(1+w_d)\sigma_d\right]\equiv S(\tau).
\end{align}
The right hand side acts as a driving force or source $S(\tau)$ for the harmonic oscillator. 
The Greens function for this differential equation is $G(\tau)=\sqrt{3}/k\ \sin(k\tau/\sqrt{3})$, such that we can formally solve the above equation (adapted from e.g. \cite{Figueroa:2012kw,Machado:2018nqk}) and find 
\begin{align}
    \epsilon^{\lim}_{ac}(k)&= \frac{1}{8} \mathcal{P}_{\delta_\gamma}(k,\tau_{\lim})+\frac{2}{3} \mathcal{P}_{v_\gamma}(k,\tau_{\lim})\\
    &=\frac{3}{8}\frac{1}{k^2}\int_{\tau_*}^{\tau_{\lim}}d\tau'\int_{\tau_*-\tau'}^{\tau_{\lim}-\tau'}d\tau''\ \cos\left(\frac{k\tau''}{\sqrt{3}}\right)\mathcal{P}_{S}(k,\tau',\tau'+\tau'')\,.\label{eq:formalEac}
\end{align}
Here $\tau_{\lim}$ is chosen large enough such that $\epsilon^{\lim}_{ac}$ has approached a quasi constant value and $\tilde\delta_\gamma\approx\delta_{\gamma}$ holds, and $\tau_*$ is the time when the fluctuations in the dark sector appear. We have furthermore introduced the unequal time correlation spectrum of the source $\mathcal{P}_{S}(k,\tau,\tau')$, defined as
\begin{align}
    \langle S(\vect{k},\tau)S^*(\vect{k}',\tau') \rangle=\frac{2\pi^2}{k^3}\mathcal{P}_S(k,\tau,\tau')\ (2\pi)^3\delta^{(3)}(\vect{k}-\vect{k}')\,.
\end{align}

To make further progress we will have to specify the characteristics of the source $S(k,\tau)$. 
Assuming that the equation of state of the dark sector $w_d$ is known then so is the time dependence of $\Omega_d\propto a^{1-3w_d}$. We will hereafter assume that the dark sector behaves radiation-like such that $\Omega_d=\Omega_{d,*}=\text{const.}$ The time dependence of $\delta_d(\tau)$ and $\sigma_d(\tau)$ is, however, more intricate and closely related to the spatial structure of the dark source. Since energy is a conserved quantity, the dynamics of its fluctuations $\delta_d(\tau)$ feature some universal properties as we will see shortly. We therefore drop the shear $\sigma_d(\tau)$ from the source term in order to simplify the discussion.

\subsection{Spatial Structure}
\label{sec:spatial_structure}

It is reasonable to assume that the mechanism that causes the fluctuations in the dark sector has an intrinsic length scale or at least a finite range of scales over which sizeable fluctuations get produced. 
We assume here that there is only one characteristic scale $k_*$ that due to causality has to lie within the horizon when the fluctuations get produced, $a_*H_*<k_*$.\footnote{Generalisation of our results is however straight forward as long as one may consider the different length scales independently.} Since there is only one characteristic scale, the fluctuations that become separated by distances greater than $1/k_*$ are uncorrelated,
\begin{align}
    \langle\delta_d(\vect{x})\delta_d(\vect{y})\rangle\approx 0\,;\qquad |\vect{x}-\vect{y}|>1/k_*\,.
\end{align}
Distributions where there is no correlation past a certain scale are commonly referred to as ``white''. For concreteness we will use
\begin{align}
    \langle\delta_d(\vect{x})\delta_d(\vect{y})\rangle=A_{\delta_d}\exp\left(-\frac{|\vect{x}-\vect{y}|^2k_*^2}{2}\right)\quad\Longrightarrow\quad \Pow_{\delta_d}(k)=A_{\delta_d}\sqrt{\frac{2}{\pi}}\frac{k^3}{k_*^3}\exp\left(-\frac{k^2}{2k_*^2}\right),\label{eq:PowAnsatz}
\end{align}
where $A_{\delta_d}$ parameterises the amplitude of the fluctuations. The common feature of white distributions in three dimensions is that their power spectrum falls off as $k^3$ in the infrared, while the UV behavior depends on the exact shape. Had we chosen a distribution with compact support in position space, the power spectrum would fall off as a power-law in the UV instead of exponentially. A power-law in the UV might lead to potentially larger signals in cases where the energy injection happens shortly before CMB emission, such that only the UV tail overlaps with the window function. The exponential fall off therefore represents a conservative choice.
The power spectrum gives the value of the unequal time correlation spectrum when one chooses both times to be the same $\Pow_{\delta_d}(k,\tau)=\Pow_{\delta_d}(k,\tau,\tau)$ and therefore gives the amplitude of the fluctuations at a given time. Since, the power spectrum falls for $k>k_*$ and the gravitational interaction for modes deeper inside the horizon is weaker, we can already anticipate that the acoustic energy becomes dominated by modes with $k\lesssim k_*$. For this reason, we only consider these modes in the following i.e. we only deal with length scales that are large enough such that there are no correlations past them.  

\subsection{Time Evolution}\label{sec:time_evol}
Here we will make the Ansatz that the energy fluctuations of the dark sector can be described as a stationary statistical process past $\tau_*$. This means that the unequal time correlation spectra can be factorized into a time autocorrelation function and a power spectrum. The power spectrum becomes constant past $\tau_*$ and the autocorrelation function $\mathcal{A}_{\delta_d}$ only depends on the difference in time
\begin{align}
\Pow_{\delta_d}(k,\tau,\tau')=\Pow_{\delta_d}(k)\mathcal{A}_{\delta_d}(k,\tau-\tau')\ \theta(\tau-\tau_*)\theta(\tau'-\tau_*)\,.
\end{align}
Let us start by considering a dark sector with relativistic dynamics. In this case, one naively expects that the only relevant time scales are $1/k_*$ and $1/k$. 
Because energy is a conserved quantity though, a change of $\delta_d$ on sub-horizon scales corresponds to a displacement of energy over a distance of $\approx1/k$. This is why the only time scale for energy fluctuations to change is given by $\approx c_{d}/k$, where $c_{d}\leq1$ is the typical velocity of energy transport in the dark sector.\footnote{Note that since the radiation dominated FRW universe possesses no time-like Killing vector field, there is no global energy conservation. On super-horizon scales modifications of energy conservation by pressure fluctuations become relevant as e.g. observed in models of cosmic seeds \cite{Durrer:2001cg,Durrer:1997te}.}

Since the energy fluctuations have this universal behavior, we limited the discussion to them and dropped the shear $\sigma_d$ from the source $S$. In general we expect the shear to be of the same size as the density fluctuations $\delta_d$ and this approximation therefore introduces an $\mathcal{O}(1)$ uncertainty.

Below we calculate the acoustic energy for two examples. In the first case the energy fluctuations exhibit a stochastic behavior and the autocorrelation therefore decays as $\mathcal{A}_{\delta_d}(k,\Delta\tau)\rightarrow0,\ |\Delta\tau|\rightarrow\infty$. For the other, we take a periodic, deterministic behavior as one expects if the dark sector comprises a fluid with waves itself.
\subsubsection*{Stochastic Source: Free Scalar Field}
For a relativistic scalar field with Gaussian fluctuations, the autocorrelation function of the energy fluctuations for $k\ll k_*$ is given by
\begin{align}
    \mathcal{A}_{\delta_d}(k,\Delta \tau)=\text{sinc}(k\,\Delta\tau)\,,
    \label{eq:autocorr_free}
\end{align}
as we show explicitly in \cref{sec:FreeScalarField}.
Since the autocorrelation decays much faster than a Hubble time if $k\gg a_{*}H_{*}$, we approximate $\epsilon_{ac}^{\lim}$ as
\begin{align}
    \epsilon^{\lim}_{ac}(k)
    &=\Omega_{d,*}^2\Pow_{\delta_d}(k)\ \frac{6}{k^2}\int_{\tau_*}^{\infty} d\tau' a^{4}(\tau')H^{4}(\tau')\int_{-\infty}^{\infty} d\tau''\ \cos\left(\frac{k\tau''}{\sqrt{3}}\right)\text{sinc}(k\tau)\\
    &=\Omega_{d,*}^2\Pow_{\delta_d}(k)\ 2\pi \left(\frac{a_{*}H_{*}}{k}\right)^3 \,,
\end{align}
where we used $a=\tau\ H_{*}a^2_{*}$ during radiation domination to solve the first integral. This estimate holds only for modes that are inside the horizon at $a_*$. The numerical results we present below suggest that for super horizon modes the efficiency of inducing acoustic waves is directly proportional to the amplitude $\Pow_{\delta_d}(k)$. We therefore use 
\begin{align}
    \epsilon^{\lim}_{ac}(k)
    &\approx\Omega_{d,*}^2\Pow_{\delta_d}(k)\ \frac{\pi}{2\pi+\left(k/a_{*}H_{*}\right)^2}\frac{1}{1+k/(2a_{*}H_{*})} \,,\label{eq:anaEst_stoch}
\end{align}
to estimate the acoustic energy for all regimes with $\mathcal{O}(1)$ accuracy. 

\subsubsection*{Deterministic Source: Fluid}

    \begin{figure}
        \centering
        \includegraphics[width= \textwidth]{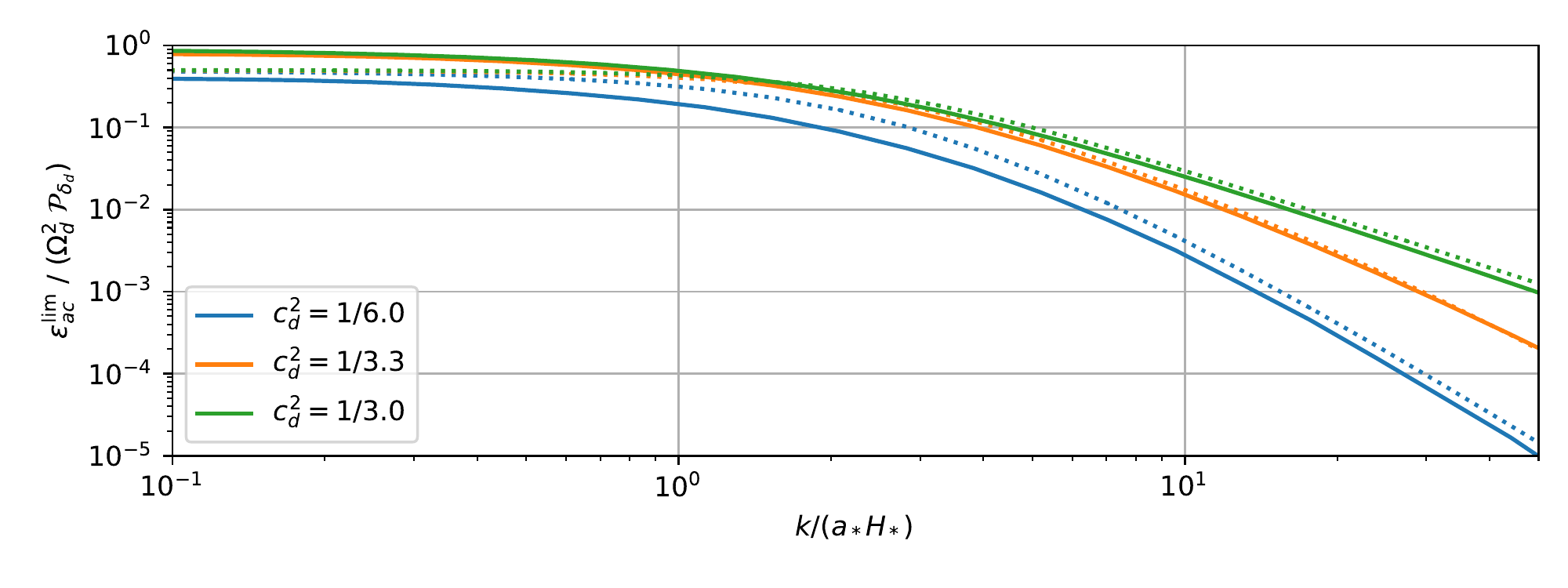}
        \vspace*{-2\baselineskip}
        \caption{
        Acoustic energy $\epsilon^{\lim}_{ac}(k)$ induced by sound waves in a dark fluid, normalized to the magnitude of the dark fluctuations $\Omega_{d,*}^2\Pow_{\delta_d}$. The efficiency is roughly constant for super-horizon modes $k< a_{*}H_{*}$, while it is suppressed for modes that are inside the horizon $k\gtrsim a_{*}H_{*}$ when the fluctuations develop at $a_{*}$. 
        Solid lines show the result from a numerical simulation including the contributions of the neutrinos and baryon-photon fluid to the gravitational potentials, while the dotted lines show the estimate $\cref{eq:anaEst_det}$. For $k\gg a_{*}H_{*}$ the suppression falls as $\propto(a_{*}H_{*}/k)^2$ in the resonant case $c_d=c_{\gamma}=1/\sqrt{3}$ (green) and as $\propto(a_{*}H_{*}/k)^4$ in the off-resonant cases, once the discrepancy in frequency becomes relevant.}
        \label{fig:ac_energy_suppression}
    \end{figure}

If the dark sector is comprised of a fluid itself with speed of sound $c_{d}$ the autocorrelation is
\begin{align}
    \mathcal{A}_{\delta_d}(k,\Delta \tau)=\cos(c_{d}k\,\Delta\tau).
\end{align}
In this case there is no sensible approximation that allows one to factorize the double integral in \cref{eq:formalEac}. Therefore we directly use the results from solving the full equations of motion numerically, including backreaction from photons and neutrinos, to discuss the behaviour. 
To do so we model the fluctuations in the dark sector as $0$ up to $\tau_*$ and as $\delta_{d}=\sin(c_{d}k(\tau-\tau_*))$ afterwards. This corresponds to $\Pow_{\delta_{d}}(k)=\langle\sin^2\rangle=1/2$. In \cref{fig:ac_energy_suppression} we show the results for $\epsilon^{\lim}_{ac}$ normalized by $\Omega_{d,*}^2\Pow_{\delta_d}$ for various dark speeds of sound. As one can see the efficiency of inducing acoustic waves takes on a constant $\mathcal{O}(1)$ value in all cases for modes outside the horizon at $a_*$.  For modes inside the horizon the efficiency falls off as $(a_*H_*/k)^2$ and as $(a_*H_*/k)^4$ once the potential offset in frequency between the driving force $c_{d}k$ and $k/\sqrt{3}$ of the driven oscillator $\delta_{\gamma}$ becomes relevant $(\sqrt{3}-c_{d})k/(a_{*}H_{*})\gtrsim1$. We therefore suggest using 
\begin{align}
    \epsilon^{\lim}_{ac}(k)
    &\approx\Omega_{d,*}^2\Pow_{\delta_d}(k)\ \frac{\pi}{2\pi+\left(k/a_{*}H_{*}\right)^2}\frac{1}{1+(1/3-c_{d}^2)\left(k/a_{*}H_{*}\right)^2} \,,\label{eq:anaEst_det}
\end{align}
which matches the numerical result up to a factor of $\lesssim 2$ for the cases shown in \cref{fig:ac_energy_suppression}.

One can interpret the results \cref{eq:anaEst_stoch} and \cref{eq:anaEst_det} as follows: Naively one expects a suppression by $(a_{*}H_{*}/k)^4$ for modes inside the horizon, $k>a_{*}H_{*}$, that stems from the potentials decaying as $\Phi,\Psi\propto(a_{*}H_{*}/k)^2$. The potentials are however approximately constant and keep driving the acoustic oscillations in the baryon-photon fluid for a whole Hubble time after $a_*$, corresponding to many oscillations $N_\text{osc}\propto k/a_{*}H_{*}\gg1$ of the mode.
The energy of a harmonic oscillator driven by a stochastic source or in resonance grows as $\propto N_\text{osc}$ or $\propto N_\text{osc}^2$ which leads to a milder suppression by $(a_{*}H_{*}/k)^3$ and $(a_{*}H_{*}/k)^2$ respectively.

\section{Detailed analysis of a simple model: $\lambda\phi^4$-Theory}
\label{sec:lambdaphi4}
To confirm the validity of our analytic estimates, we now consider a toy model for which a full numerical treatment is feasible. 
We have chosen a model consisting of two real scalar fields $\phi,\psi$ with the potential 
\begin{align}
    V(\phi,\psi)=\frac{1}{4}\lambda\phi^4+\frac{1}{2}g^2\phi^2\psi^2\,.
\end{align}
This model has been studied in great detail in the context of preheating e.g. \cite{Greene:1997fu,Kofman:1994rk,Shtanov:1994ce,Kofman:1997yn,Figueroa:2016wxr,Figueroa:2017vfa}.  We consider $\langle\phi\rangle=\phi_i\ll\mPl$ and $\langle\psi\rangle=\langle\dot\phi\rangle=\langle\dot\psi\rangle=0$ as the initial conditions. In this case, the energy density of the fields is always subdominant $\Omega_{d}\ll1$ in contrast to the preheating scenario where the field $\phi$ with $\phi_i>\mPl$ initially drives inflation. 

In addition to deriving all the parameters needed to estimate the acoustic energy analytically as described above, we solve the dynamics of the model from first principles using a lattice simulation, which allows us to extract $\delta_d(k,\tau)$ and $\sigma_d(k,\tau)$. Using these we solve for  $\delta_\gamma(k,\tau)$ and arrive at the acoustic energy. We recommend the reader only interested in the application of the analytic estimates to skip forward to the next section.

We follow \cite{Figueroa:2016wxr,Figueroa:2017vfa} in our analysis and define the angular frequency $\omega_*=\sqrt{\lambda}\phi_i$ giving the typical curvature of the potential as well as the resonance parameter $q=g^2/\lambda$. For simplicity, we fix the latter at $q=1$ in this work. Furthermore, we assume that both $\phi$ and $\psi$ possess Gaussian fluctuations originating from inflation that are frozen before a mode enters the horizon
\begin{align}
    \mathcal{P}_{\phi,\psi}=\left(\frac{H_{I}}{2\pi}\right)^2;\qquad \mathcal{P}_{\dot\phi,\dot\psi}\approx0\,,
\end{align}
where $H_{I}\gg \omega_*$ denotes the Hubble parameter during inflation. We assume that after inflation the universe reheats and undergoes the same evolution as in the standard $\Lambda$CDM case, with our dark sector acting as a purely gravitationally coupled spectator. At the times relevant for generating $\mu$-distortions the universe is still radiation dominated.

\subsection{Analytic Estimate}
\label{sec:lphi4_ana}
The field $\phi$ starts to oscillate once the Hubble rate drops to $H_\text{osc}=\omega_*$. The energy in the field $\phi$ is initially $\omega^2_*\phi_{i}^2/4$ and dominates the dark sector such that one can estimate $\Omega_{d,osc}\propto \left({\phi_{i}}/{\mPl}\right)^2$.
Past $a_\text{osc}$ the dark sector behaves like radiation such that $\Omega_{d}\approx\text{const}$. To go beyond an order of magnitude estimate, one has to solve the equation of motion for the homogeneous component of $\phi$ and finds 
\begin{align}
    \Omega_{d,osc}\simeq0.2\left(\frac{\phi_{i}}{\mPl}\right)^2.\label{eq:Omega_lphi4}
\end{align}
The oscillations of $\phi$ lead to a time-dependent effective mass of the field $\psi$, which cause its fluctuations to grow exponentially. As shown in e.g. \cite{Greene:1997fu} the equation of motion for the Fourier modes of the field $\psi$ can be recast into the \textit{Lam\'e} equation. From the corresponding instability chart, one can read off that the modes with $k\lesssim\omega_*a_\text{osc}$ experience exponential growth. The mode growing the fastest is $k_{*}\approx \omega_* a_\text{osc}/\sqrt{2}$ with its energy density growing as $\propto \exp(0.3\, \omega_* a_\text{osc}\tau )$. The energy in the fluctuations is initially $\approx\omega^2_*H_{I}^2/(2\pi)^2$ while the energy in the homogeneous $\phi$ field is $\approx\omega^2_*\phi_{i}^2/4$.  Due to the exponential growth this difference is overcome around
\begin{align}
    a_{*}\approx a_\text{osc} \frac{2}{0.3}\log\left(\frac{\pi\phi_{i}}{H_{I}}\right)\,.\label{eq:astar}
\end{align}
At this point, the energy in the fluctuations starts to dominate, causing the energy density to become fully inhomogeneous in line with the definition of $a_*$ in the previous chapters. This allows us to calculate $a_{*}H_{*}= a_\text{osc}\omega_*\cdot a_\text{osc}/a_*$. For the simulations presented in the following we fixed $H_{I}/\phi_{i}=10^{-4}$, which gives $a_*\approx 70\ a_\text{osc}$ and $a_{*}H_{*}\approx a_\text{osc}\omega_*/70$.

Once the fluctuations dominate, the energy gets split between the two fields and their respective kinetic and gradient contributions. If the system virializes quickly, the energy will be distributed evenly between the four, and there will be no correlations between them. If each separate contribution has $\mathcal{O}(1)$ fluctuations we find $\langle\delta_{d}^2\rangle=A_{\delta_d}=1/4$, such that we can use \cref{eq:PowAnsatz} to estimate $\Pow_{\delta_d}(k)$.

As a final step, we need to make an assumption about the temporal behavior of $\delta_d(k,\tau)$. Similar to the case of the free scalar field, the energy fluctuations are due to the random interference of the field modes. If anything, one expects the potentially turbulent interaction of the field modes at $a_*$ to lead to a faster decrease in the autocorrelation function. We, therefore, use \cref{eq:anaEst_stoch} with the parameters derived above to analytically estimate the induced acoustic energy.

\subsection{Numerical Treatment}
Using \verb|CosmoLattice| \cite{Figueroa:2020rrl,Figueroa:2021yhd} we solve the full equations of motion of the interacting $\phi$ and $\psi$-field on a discretized space-time using a second-order Velocity Verlet algorithm (equivalent to using a leapfrog algorithm). The evolution of the background metric is set to behave like a radiation dominated universe, independently of the dark sector. We simulate a box with $N=1024$ sites along each spatial direction with a comoving length of $L=2\pi a_\text{osc}/(0.015\ \omega_*)$ and periodic boundary conditions. The fields in this box are evolved by time steps of $d\tau=0.05\ a_\text{osc}/\omega_*$. While this choice compromises between covering the dynamics close to the horizon at $a_*$ and resolving the UV dynamics, once the system becomes fully non-perturbative, we ran simulations with higher spatial resolution and smaller time steps to ensure that none of our results are affected by the poor UV resolution of the run presented here. 

We start the simulation at $a_i=a_\text{osc}/10$ and use the initial conditions given above. We cut the inflationary spectrum off for $k>1.3\ a_\text{osc}\omega_*$ to cover the full instability band in $\psi$, while at the same time only including modes with $k\ll aH|_{a=a_i}=10\ a_\text{osc}\omega_*$ such that $\mathcal{P}_{\dot\phi,\dot\psi}\approx0$ holds. After fixing $q=1$ and $H_{I}/\phi_{i}=10^{-4}$, the only remaining free parameters are $\phi_i$ and $\lambda$, or equivalently $\phi_i$ and $\omega_*$. The dependence on these two is however fully covered by the scaling relations discussed above, with $\phi_i$ controlling $\Omega_d$ and $\omega_*$ the typical momentum scale $k_{*}$. We keep these relations explicit when showing our results below.

 We modified \verb|CosmoLattice| to calculate and output $\delta_{d}(k,\tau)$ and $\tilde\sigma_{d}(k,\tau)$ in time-intervals of $\Delta\tau=0.5\ a_\text{osc}/\omega_*$. Here we defined
\begin{align}
    \tilde\sigma_{d}(k,\tau)=(1+w)\sigma_{d}(k,\tau)\,.
\end{align}
\VerbatimFootnotes
This is more convenient for numerics, since it does not require knowledge of the average pressure in the dark sector. The details of how we calculated these quantities can be found in \cref{sec:NumericsPhi4}. To obtain the power spectra of these quantities, we group them in radial bins of width $k_{IR}=0.015\ \omega_*$ and average over them. To keep the computational cost and required storage down, we limit ourselves to 70 bins that are spaced out linearly at low $k$ and logarithmically at high $k$ and only use up to 1000 modes per bin.\footnote{Our method is equivalent to the type II, version 1 powerspectrum from the \verb|CosmoLattice| technical note \cite{CosmoLattice:technical}, except for limiting the number of modes per bin. \verb|CosmoLattice| includes modes up to $\sqrt{3}/2\ Nk_{IR}$, while we limited ourselfs to $1/2\ Nk_{IR}$, which explains why the spectra calculated by \verb|CosmoLattice| directly extend to slightly higher momenta than the once calculated by our methods in \cref{fig:spectra_lattice}.} By interpolating between the saved values of $\delta_{d}(k,\tau)$ and $\tilde\sigma_{d}(k,\tau)$ as well as $\Omega_d(\tau)$ one can solve for the perturbations in the visible sector for each single mode(see \cref{sec:TCA_nuDyn}). With this approach one doesn't have to make any assumptions about the time evolution like we did in \cref{sec:time_evol}. To obtain the induced acoustic energy $\epsilon^{\lim}_{ac}$ or $\mu$-distortion we then take the power spectra of $\Pow_{\delta_\gamma}(k,\tau)$ and $\Pow_{v_\gamma}(k,\tau)$  by averaging over the modes in one bin again and use \cref{eq:ac_energy,eq:ac_power}.

\subsection{Numerical Results}\label{sec:num_results}
\begin{figure}
        \centering
        \includegraphics[width= \textwidth]{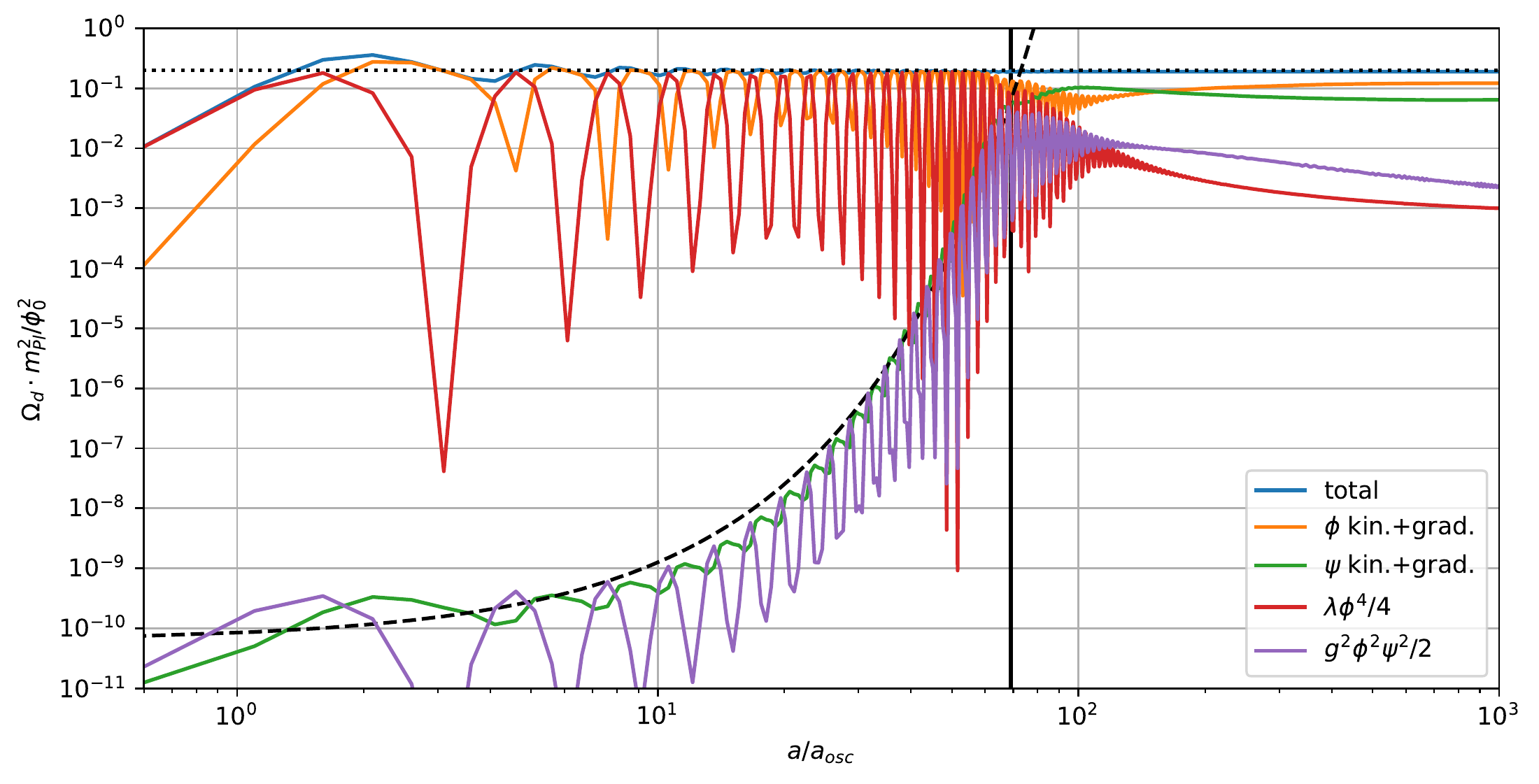}
        \vspace*{-1.5\baselineskip}
        \caption{Evolution of energy components in the dark sector. Around $a_\text{osc}$ the $\phi$-field starts oscillating and the energy initially stored in the quartic potential (red) starts going back and forth between the potential and kinetic energy (orange). The dark sector transitions from vacuum to radiation like scaling and its total energy (blue) asymptotes to the value given in \cref{eq:Omega_lphi4} (dotted, black). The instabilities induced by the coupling (purple) in $\psi$ lead to its energy (green) growing exponentially past $a_\text{osc}$. The mode functions of $\psi$ can be approximated by solutions to the \textit{Lam\'e} equation, leading to the estimate given by the dashed black line. The horizontal black line marks $a_*$, the time when the energy in the homogeneous $\phi$ field equals the energy in inhomogenities of $\psi$ as estimated in \cref{eq:astar}. At this point the perturbative treatment breaks down, making the lattice analysis necessary. Past this point the majority of energy is stored in fluctuations of $\phi$ and $\psi$.}
        \label{fig:energies_evolution}
\end{figure}

\begin{figure}
        \centering
        \includegraphics[width= \columnwidth]{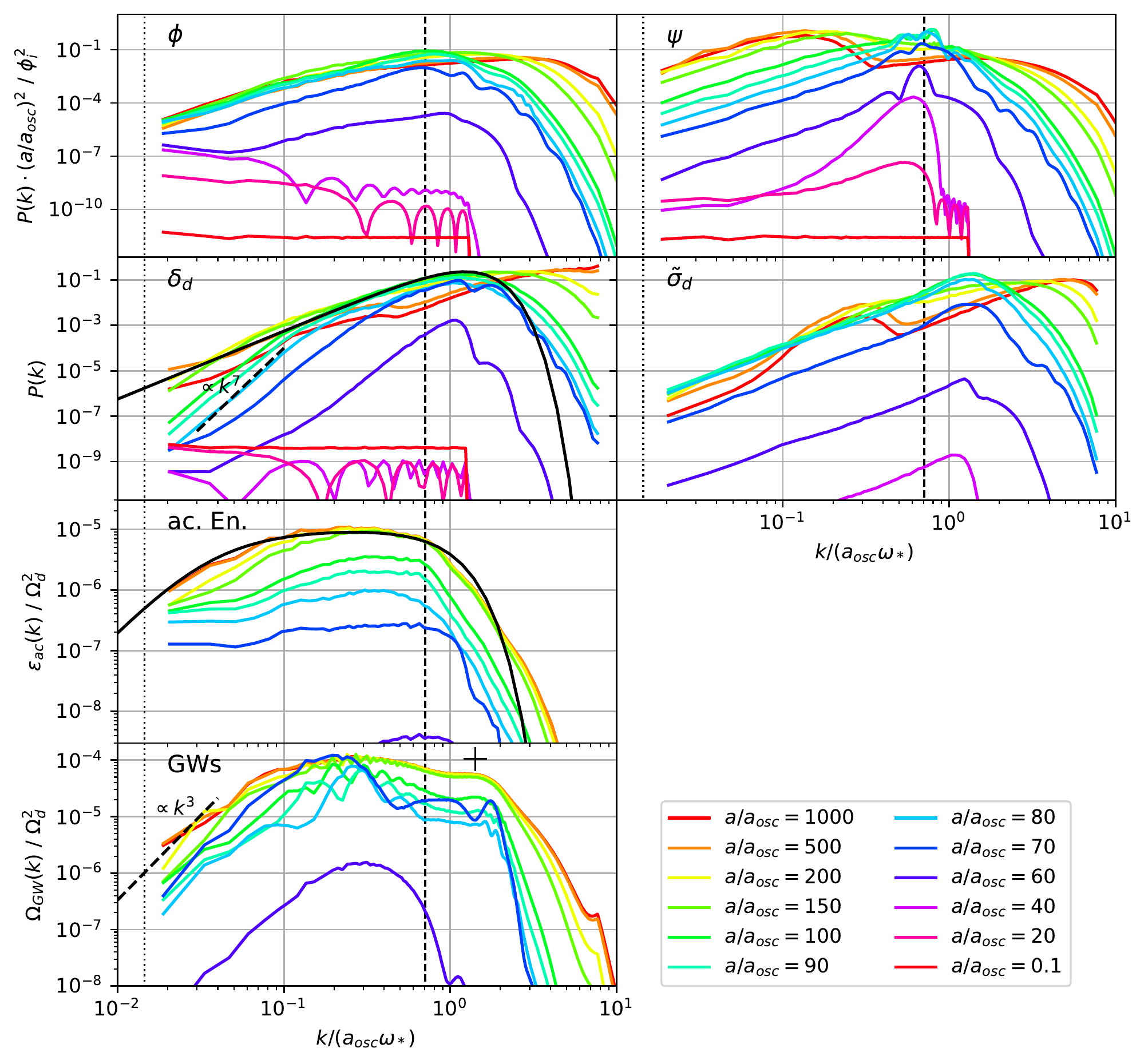}
        \vspace*{-1.5\baselineskip}
        \caption{Evolution of power spectra of the $\phi$ and $\psi$-field  (top row) as well as energy fluctuations $\delta_d$ and shear $\tilde\sigma_d$ in the dark sector (second row). In the third row we show the induced acoustic energy in the baryon-photon fluid through gravitational coupling. See \cref{sec:num_results} for discussion. In the bottom row we furthermore show the resulting spectrum of gravitational waves as discussed in \cref{sec:GWs_lphi4}. The vertical, black, dotted line marks the horizon scale at $a_*$, $k\approx a_\text{osc}\omega_*/70$, and the dashed line gives our estimate for the fastest growing mode in $\psi$, $k_*\approx a_\text{osc}\omega_*/\sqrt{2}$. The thick, black, solid and dashed lines show the analytic estimates discussed in the text. Both solid lines feature the exponential decay in the UV as a result of our conservative choice in \cref{sec:spatial_structure} with the actual UV tail falling off more gradually.}
        \label{fig:spectra_lattice}
\end{figure}
In \cref{fig:energies_evolution} we show the evolution of various energy components in the dark sector. Up to $a_\text{osc}$ the energy is almost exclusively stored in the quartic potential while the dynamics of $\phi$ remains overdamped by Hubble friction. Around $a_\text{osc}$, $\phi$ starts to oscillate and the energy in the dark sector red-shifts like radiation, resulting in $\Omega_d$ taking on the value given in \cref{eq:Omega_lphi4}. 
The fluctuations stored in $\psi$ are subdominant around $a_\text{osc}$ but start growing exponentially due to the instability caused by the coupling to the oscillating $\phi$-field. The black, dashed line shows the analytic estimate obtained by looking up the growth coefficient in the instability chart of the \textit{Lam\'e} equation
\begin{align}
    \Omega_{\psi}(\tau)\approx\frac{1}{3}\left(\frac{H_{I}}{2\pi\mPl}\right)^2\exp(0.3\, \omega_* a_\text{osc}(\tau-\tau_\text{osc} ))\,.
\end{align}
Once the energy in fluctuations of $\psi$ catches up to the energy in the $\phi$-field, there is a back-reaction that decreases the amplitude of oscillations of the homogeneous part of $\phi$ while at the same time introducing sizeable inhomogeneities in $\phi$. Shortly after $a_*$ the energy becomes dominated by the gradient and kinetic terms, corresponding to fluctuations of $\phi$ and $\psi$, with the potential energy decreasing.

The evolution of the fluctuations in the fields can be seen directly from the top row of \cref{fig:spectra_lattice}. While the fluctuations in the $\phi$ field only oscillate as they enter the horizon, leading to the fringe pattern, the fluctuations in the $\psi$ field grow exponentially in the instability band $k\lesssim a_\text{osc}\omega_*$ with the modes around $k_*$, marked by the vertical dashed line, growing the fastest. At $a/a_\text{osc}=60$, as the system approaches $a_*$, we can see first signs of a back-reaction in the form of additional induced fluctuations in $\phi$. Past $a_*/a_\text{osc}=70$ both spectra feature a primary peak that keeps moving to higher $k$ as time progresses. This can be understood as the onset of thermalization as $\phi$ and $\psi$ particles/waves scatter off one another \cite{Boyanovsky:2003tc,Micha:2004bv,Destri:2004ck}. Somewhat surprisingly there forms a secondary peak in the spectrum of $\psi$ around $k\approx 0.1\ a_\text{osc}\omega_*$. We can only speculate that this might be the result of the homogeneous part of $\phi$ being damped and the instability band therefore moving to lower $k$. 

In the second row of \cref{fig:spectra_lattice} we show the evolution of power spectra of the density fluctuations in the dark sector as well as the shear. Initially the spectrum of density fluctuations is due to the interference of fluctuations of $\phi$ with its homogeneous component that still dominates the energy. The spectrum is therefore also initially flat, as expected, and shows an oscillatory pattern similar to that of the modes entering the horizon. Around $a_{*}\approx 70\, a_\text{osc}$, the fluctuations in the energy density are well described by the analytic estimate \cref{eq:PowAnsatz} with the parameters derived in \cref{sec:lphi4_ana} (straight, black line). Our estimate describes the energy fluctuations well for the Hubble time following $a_{*}$, which is when we expect most of the acoustic energy in the baryon-photon fluid to be sourced. At later times the peak moves to higher $k$ as a result of the scattering processes discussed above. The evolution of the shear is similar although it develops a much more pronounced secondary peak at late times.

\begin{figure}
        \centering
        \includegraphics[width= \columnwidth]{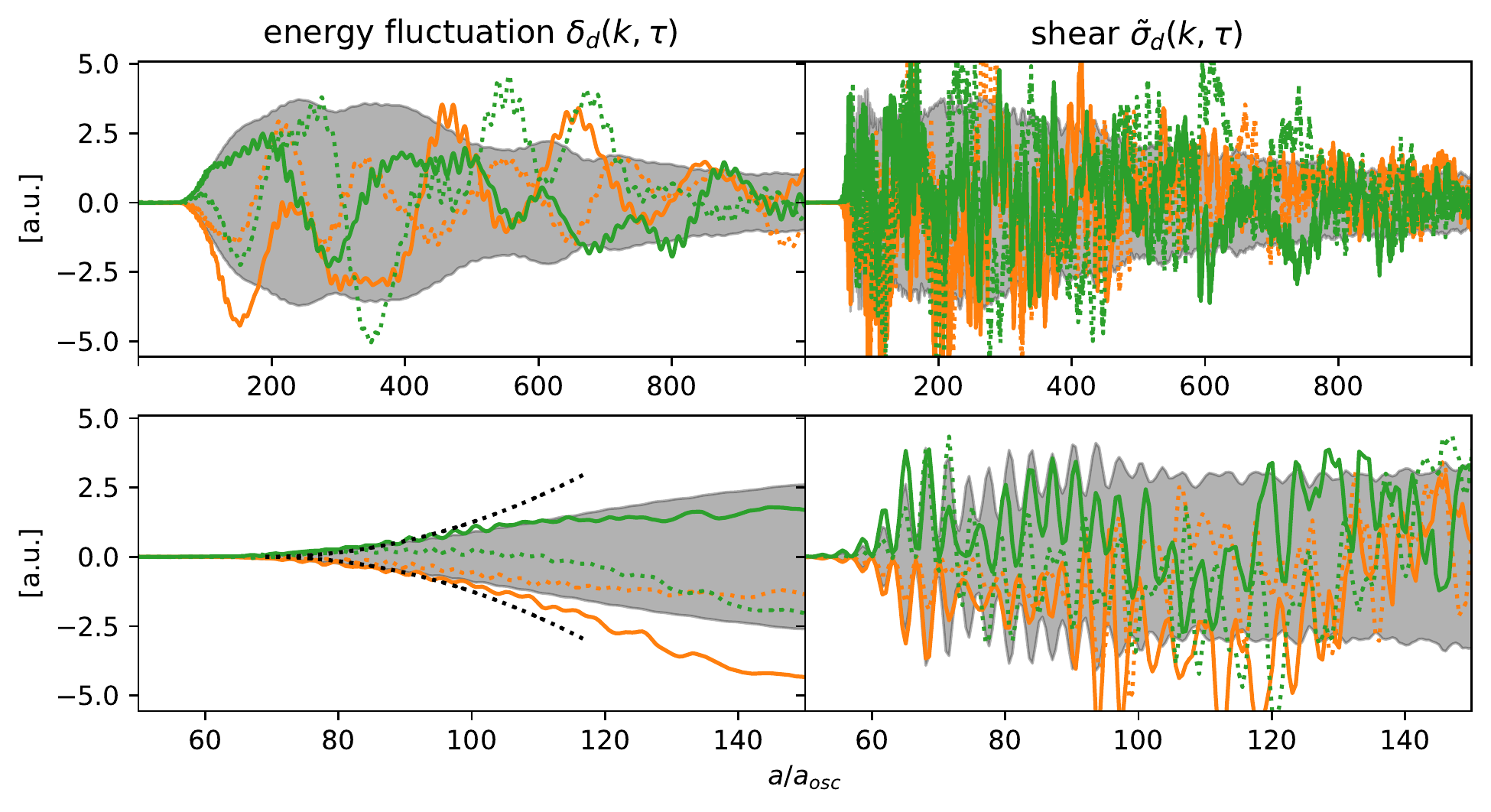}
        \vspace*{-1.5\baselineskip}
\caption{Top row: Evolution of the energy fluctuation $\delta_d(k,\tau)$ and the shear $\tilde\sigma_d(k,\tau)$ for two modes (orange and green) in the infrared tail of the spectrum ($k=0.05\ a_\text{osc}\omega_*$). The straight and dotted line give the real and imaginary part respectively. The gray envelope indicates the evolution of the power spectrum $\propto \sqrt{\Pow(k,\tau)}$ as calculated by averaging the amplitude of all mode-functions in the respective bin. Bottom row: Same as above, zoomed in on the Hubble time past $a_*\approx 70\ a_\text{osc}$. The black dotted lines indicate the amplitude of the energy fluctuations growing $\propto [k(\tau-\tau_*)]^2$ initially.}
        \label{fig:fluctuation_evolution}
\end{figure}

In \cref{fig:fluctuation_evolution} we show the evolution of the energy fluctuation $\delta_d(k,\tau)$ and the shear $\tilde\sigma_d(k,\tau)$ for two modes in the infrared tail of the spectrum. We furthermore show the average amplitude of modes in the respective bin. As there is no clear pattern visible between the two different realisations for the same $k$, a stochastic description seems to be appropriate. In line with our discussion in \cref{sec:time_evol}, we want to furthermore stress the difference in the evolution of $\delta_d(k,\tau)$ and $\tilde\sigma_d(k,\tau)$ around $a_{*}$. The shear shows $\mathcal{O}(1)$ variations on time scales $\Delta a/a_\text{osc}\approx 1 \approx k_*\Delta\tau$ that can be related to the characteristic scale $\Delta\tau\approx 1/k_*$. The energy fluctuations on the otherhand only grow as $\propto [k(\tau-\tau_*)]^2$ and take on their late time amplitude after $\Delta \tau\approx 1/k$. This can also be seen from \cref{fig:spectra_lattice} where the infrared tail of the power spectrum of the energy fluctuations is given as $\propto k^3\cdot [k(\tau-\tau_*)]^4\propto k^7$ at the times $a/a_\text{osc}=80-100$ shortly after $a_*$ before asymptoting to the final $\propto k^3$. This behavior can be understood as the energy density is conserved on sub-horizon scales and we therefore have $\dot\rho(k,\tau)=ik j_{\rho}(k,\tau)$, which leads to $\rho(k,\tau)\propto k(\tau-\tau_*)$ assuming that the corresponding current $j_{\rho}(k,\tau)$ jumps to its final amplitude around $a_*$. Since the current $j_{\rho}(k,\tau)$ is however the momentum density and itself conserved, one finds $\rho(k,\tau)\propto [k(\tau-\tau_*)]^2$.

\begin{figure}
        \centering
        \includegraphics[width= \textwidth]{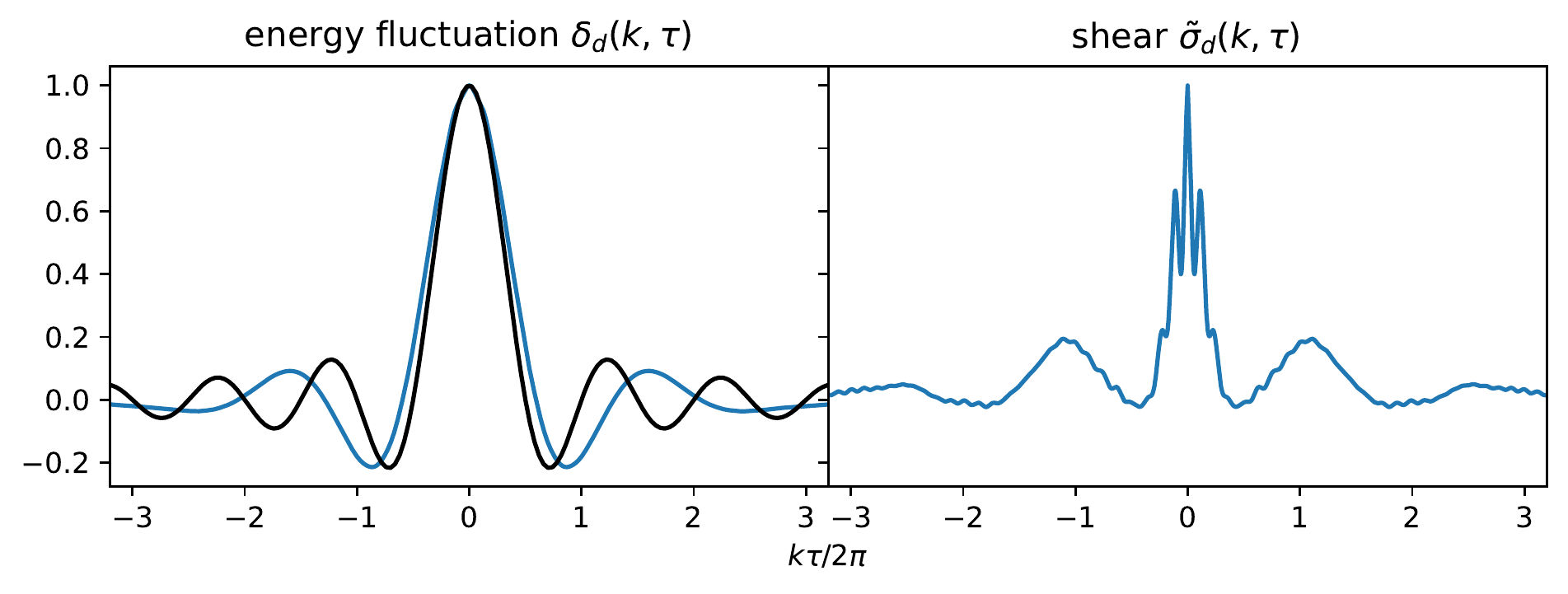}
        \vspace*{-1.5\baselineskip}
        \caption{Autocorrelation of the energy fluctuation $\delta_d(k,\tau)$ and the shear $\tilde\sigma_d(k,\tau)$ for same momentum  as in \cref{fig:fluctuation_evolution} ($k=0.05\ a_\text{osc}\omega_*$). For the energy fluctuation we show for comparison the $\text{sinc}$ that we find analytically in \cref{sec:FreeScalarField} for a free scalar field (black).}
        \label{fig:autocorrelation}
\end{figure}

In \cref{fig:autocorrelation} we show the autocorrelation function of $\delta_d(k,\tau)$ and  $\tilde\sigma_d(k,\tau)$ for the same values of $k$, calculated by averaging over the modes in the respective bin and times between $a/a_\text{osc}=200-1000$. For comparison we also show the analytic approximation~\cref{eq:autocorr_free} for the energy fluctuations of a free scalar field (\cref{sec:FreeScalarField}). As one can see there is good qualitative agreement in that they both have a central peak of width $\approx 1/k$. 
As argued in \cref{sec:time_evol} this is also expected from energy conservation. Finding these two features makes us confident that our lattice version of the energy density indeed resembles the continuum one.\footnote{We first tried to do this analysis for an axion coupled to a vector, using the same code as in \cite{Ratzinger:2020oct}. This model and its lattice implementation is more complicated since it involves vectors. We were not able to construct an energy density on the lattice that showed these characteristics of energy conservation without decreasing the time step of the simulation by a lot, making the simulation unfeasible. We leave a systematic investigation of this issue for future work and recommend checking these features when running similar simulations.}
We also find qualitative agreement for the autocorrelation of the shear from the lattice simulation and the free scalar field. Both have features on small time scales related to the peak momentum $k_*$ and on time scales related to $k$. It should however be mentioned that the autocorrelation function of the shear varies much more when varying the momentum $k$.

\begin{figure}
        \centering
        \includegraphics[width= \textwidth]{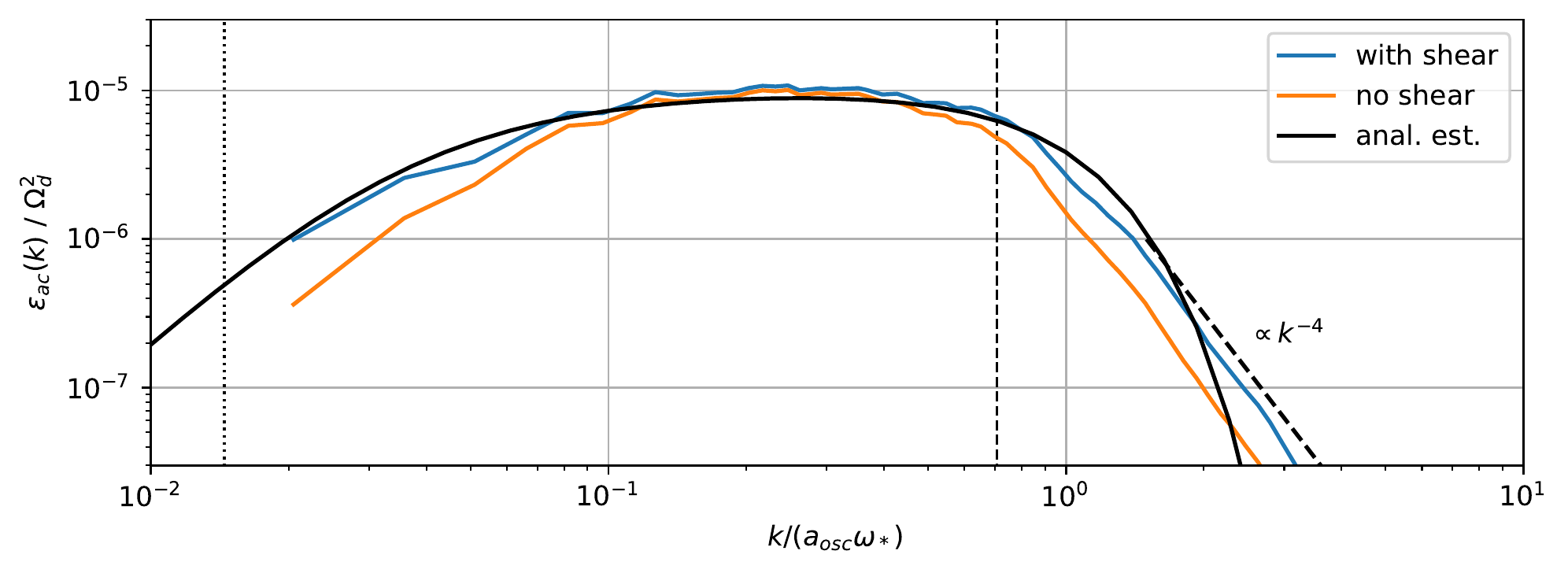}
        \vspace*{-1.5\baselineskip}
        \caption{Close-up of the acoustic energy spectrum at the final time of the simulation. The blue line represents the physical result, while the orange line stems from a simulation in which the shear of the dark sector was neglected. The straight black line gives the analytical estimate, with the black dashed line indicating a $k^{-4}$ power law that seems more appropriate to describe the UV tail than the exponential decay of the analytic estimate.  The vertical, black dotted and dashed line give the horizon at $a_*$ and the estimate of the fastest growing mode in the $\psi$-field, $k_*$.}
        \label{fig:ac_en_spectrum}
\end{figure}

Given the evolution of fluctuations in the dark sector as shown in \cref{fig:fluctuation_evolution}, we can numerically solve the equations for fluctuations in the baryon-photon fluid on a mode per mode basis. This allows us to calculate the acoustic energy and the result is shown in the third row of \cref{fig:spectra_lattice} and in  \cref{fig:ac_en_spectrum}. From the time evolution shown in \cref{fig:spectra_lattice} we see that
the majority of acoustic energy is induced in the Hubble time after $a_*$ ($a/a_\text{osc}\approx 70-150$) and the energy becomes constant shortly after. Our analytic estimate in \cref{eq:anaEst_stoch} with the parameters derived in \cref{sec:lphi4_ana}, shown in black, accurately estimates the main features of this final spectrum: A steep fall off for modes larger than the peak momentum $k_*$, a flat plateau for momenta between $k_*$ and the horizon at $a_*$ (vertical, dotted, black line) and a $k^3$ infrared tail for momenta outside the horizon at $a_{*}$. Unfortunately our simulation does not properly cover super-horizon scales, but from what we can see the spectrum becomes steeper at the horizon in good agreement with our estimate. 

In \cref{fig:ac_en_spectrum} we show a close-up of the acoustic energy spectrum at the final time of the simulation. We furthermore show the result of a calculation in which we neglected the shear of the dark sector when solving for the perturbations in the baryon-photon fluid. We find that both calculations as well as the analytic estimate agree to within $\approx 20\%$ in the plateau region. Neglecting the shear, however, results in underestimating the acoustic energy by a factor of $\approx 2-3$ in the IR and UV tail of the spectrum. Furthermore we find that in the UV the spectrum falls of as $k^{-4}$ rather than the exponential suppression suggested by the analytic estimate.

\section{Application to sources of GWs}
\label{sec:GW_sources}
Our mechanism of sourcing $\mu$-distortions is only efficient if the dark sector features a sizeable amount of energy $\Omega_{d}\lesssim 0.04$ that has $\mathcal{O}(1)$ perturbations on scales close to the horizon. The upper bound on the energy here stems from the current $\Neff$ bound discussed below. If the dynamics of the dark sector are furthermore relativistic, they efficiently produce gravitational waves. Such models have received a large amount of attention recently due to the detection by LIGO \cite{LIGOScientific:2016aoc} and hints stemming from pulsar timing arrays (PTAs) \cite{NANOGrav:2020bcs,Goncharov:2021oub,Chen:2021rqp,Antoniadis:2022pcn}. In the following, we want to compare the reach of searches for GWs and spectral distortions for some of these models.

Before doing so, let us clarify that the inverse statement, i.e. that all sources of primordial GWs feature large perturbations close to the horizon scale, is not necessarily true. A counter example would be strings originating from a broken local $U(1)$ gauge symmetry. In this case, GWs are mainly produced from the tiniest string loops that only populate a small fraction of the Hubble volume at a given time. Therefore the source is point-like, whereas for the dark sectors we are considering, GWs get sourced over the whole Hubble volume.  

For these sectors the resulting GW spectrum is peaked around the characteristic scale $k_*$ and the peak amplitude can schematically be parameterized as \cite{Giblin:2014gra,Buchmuller:2013lra,Hindmarsh:2015qta}
\begin{align}
    \Omega_\text{GW,peak}\propto \Omega_{d}^2 \cdot \left(\frac{a_{*}H_{*}}{k_*}\right)^{\alpha}\cdot \Pow_{\delta_d}^2(k_{*})\,.\label{eq:roughEst_GW}
\end{align}
Just like the induced acoustic energy, the energy in GWs is suppressed by $\Omega_{d}^2$. The suppression originating from the characteristic scale being inside the horizon potentially differs though from Equations \ref{eq:anaEst_stoch} and \ref{eq:anaEst_det}, with the model dependent power $\alpha$ taking the values 1 and 2. If the new physics comprising the dark sector does not feature spin-2 degrees of freedom, gravitational waves can only get sourced in second-order processes resulting in the suppression by $\Pow_{\delta_d}^2\lesssim 1$. Further suppression of the GW signal occurs if the dynamics are non-relativistic, but we are not considering this case below.

\subsection*{Experiments and Cosmological Bounds}

At the lowest frequencies and correspondingly largest scales, the amount of GWs becomes limited by the non-detection of $B$-modes in the CMB polarization by Planck+BICEP2+Keck \cite{Clarke:2020bil}, and we show the resulting limit in cyan in the following plots. Furthermore, the gravitationally induced scalar fluctuations would alter the resulting CMB perturbation pattern. In our age of precision cosmology, deriving such bounds is done by refitting the angular perturbations from scratch. However, such an analysis is beyond this work. Instead, we take inspiration from searches of symmetry-breaking relics carried out in e.g \cite{Planck:2013mgr,Charnock:2016nzm,Lopez-Eiguren:2017dmc}. They found that the fraction in the angular power spectrum stemming from the new physics is limited to a couple of percent over a wide range of angular scales $l$. To visualize the remaining uncertainty, we show an aggressive bound, limiting the amount of induced fluctuations to 2\% of the inflationary ones up to the CMB pivot scale, $\Pow_{\delta_\gamma}<0.02\,\Pow_{\delta_\gamma}^\text{inf}$ for $k<0.05\,\text{Mpc}^{-1}$, as well as a more conservative bound corresponding to 10\% out to scales of $k<0.005\,\text{Mpc}^{-1}$. Here $\Pow_{\delta_\gamma}^\text{inf}\approx 16\cdot\Pow_{\xi}\approx32\times10^{-9}$ denotes the amount of inflationary fluctuations inferred form the Planck 2018 dataset \cite{Planck:2018vyg}. We calculate $\Pow_{\delta_\gamma}\approx 4 \epsilon_{ac}$ using the formulas given in \cref{sec:sound_energy}. The resulting bound is shown in red.

At smaller scales, we use the results from \cref{sec:source_mu} and \cref{sec:sound_energy} to calculate the $\mu$-distortions resulting from induced acoustic waves and show the results in green. We furthermore calculate the $\mu$-distortions stemming from the interaction of the GWs emitted by the dark sector with the baryon-photon fluid \cite{Ota:2014hha,Chluba:2014qia,Kite:2020uix}. 
We show the resulting bound in pink. The actual observable distortion would perhaps be the sum of these two effects, but we show them separately to highlight the magnitude of each source of distortion. As thresholds for the detection of a $\mu$-distortion we consider the existing bound from COBE/FIRAS $\mu<4.7\times 10^{-5}$ at 95\% confidence level \cite{Fixsen:1996nj,Bianchini:2022dqh} as well as the sensitivity of the future missions PIXIE $\mu\lesssim 3\times 10^{-8}$ \cite{Kogut:2011xw} and Voyage2050 $\mu\lesssim 1.9 \times 10^{-9}$ \cite{Chluba:2019nxa}. 

At even shorter scales, we fit the GW spectrum to the to-be-confirmed detection by pulsar timing arrays \cite{NANOGrav:2020bcs,Goncharov:2021oub,Chen:2021rqp,Antoniadis:2022pcn} using the first 5 frequency bins from the NANOGrav 12.5yr dataset \cite{NANOGrav:2020bcs} and the hierarchical method proposed in \cite{Ratzinger:2020koh,Moore:2021ibq}. Direct fits including modeling of pulsar noise have been carried out for a number of the models discussed below \cite{Ferreira:2022zzo,Chen:2022azo,NANOGrav:2021flc,Xue:2021gyq} and the results largely agree with the hierarchical method.  We show the resulting $2\sigma$ region of the fit as an orange area. Furthermore, we show the reach of the planned square kilometer (SKA) array after taking data for 20 years  \cite{Janssen:2014dka,Weltman:2018zrl,Breitbach:2018ddu} as an orange line.

Since we consider dark sectors with relativistic dynamics, they will inadvertently act as a form of radiation not interacting with the baryon-photon fluid and therefore contribute to the effective number of neutrinos $\Neff$. At recombination its contribution is given as
\begin{equation}\label{eq:Neff}
\Delta N_{\rm eff} = \frac{8}{7}\left(\frac{11}{4}\right)^{\frac{4}{3}} \frac{\rho_{d}}{\rho_{\gamma}}\bigg|_{T=T_{\rm rec}} \,.
\end{equation}
The Planck 2018 dataset constrains $\Delta N_{\rm eff} < 0.3$ at 95\% confidence level~\cite{Planck:2018vyg} and the next generation of ground-based telescopes (CMB Stage-4) is expect to achieve a sensitivity of $\Delta N_{\rm eff} < 0.03$~\cite{CMB-S4:2016ple}, which we show as a gray surface and line respectively.

\subsection{$\lambda\phi^4$-Theory}
\label{sec:GWs_lphi4}
Let's start with the model we already considered in great detail in \cref{sec:lambdaphi4}. In \cref{fig:spectra_lattice}, we show in the bottom row the evolution of the energy density spectrum of gravitational waves. As one can see, similar to the acoustic energy, the energy in gravitational waves is sourced in the Hubble time following the back-reaction of the $\psi$-field on $\phi$, $a/a_\text{osc}\approx 70-150$. One might therefore try to estimate the peak amplitude with \cref{eq:roughEst_GW} and $\alpha=2$, which has been observed to give a decent estimate for similar models. Very roughly one can set $\Pow_{\delta_d}(k_*)\approx 1$. It has also been observed that the peak of the GW spectrum lies typically about a factor $2$ higher than the characteristic scale of the source $k_\text{peak}\approx 2k_*$ (see e.g. \cite{Machado:2018nqk}). We show the resulting estimate as a black cross in \cref{fig:spectra_lattice}. Somewhat surprisingly, the actual peak of the GW spectrum lies a factor $\approx4$ below the characteristic scale $k_*$, and the estimate only corresponds to a secondary peak at higher $k$. The peak amplitude, however, is estimated to be within a factor of 2. In \cref{fig:spectra_lattice} we have furthermore indicated the $k^3$ power law that one expects for $\Omega_\text{GW}(k)$ for scales outside the horizon at $a_*$~\cite{Caprini:2009fx,Caprini:2018mtu} (dotted, black line). 

\begin{figure}
        \centering
        \includegraphics[width= 0.8\textwidth]{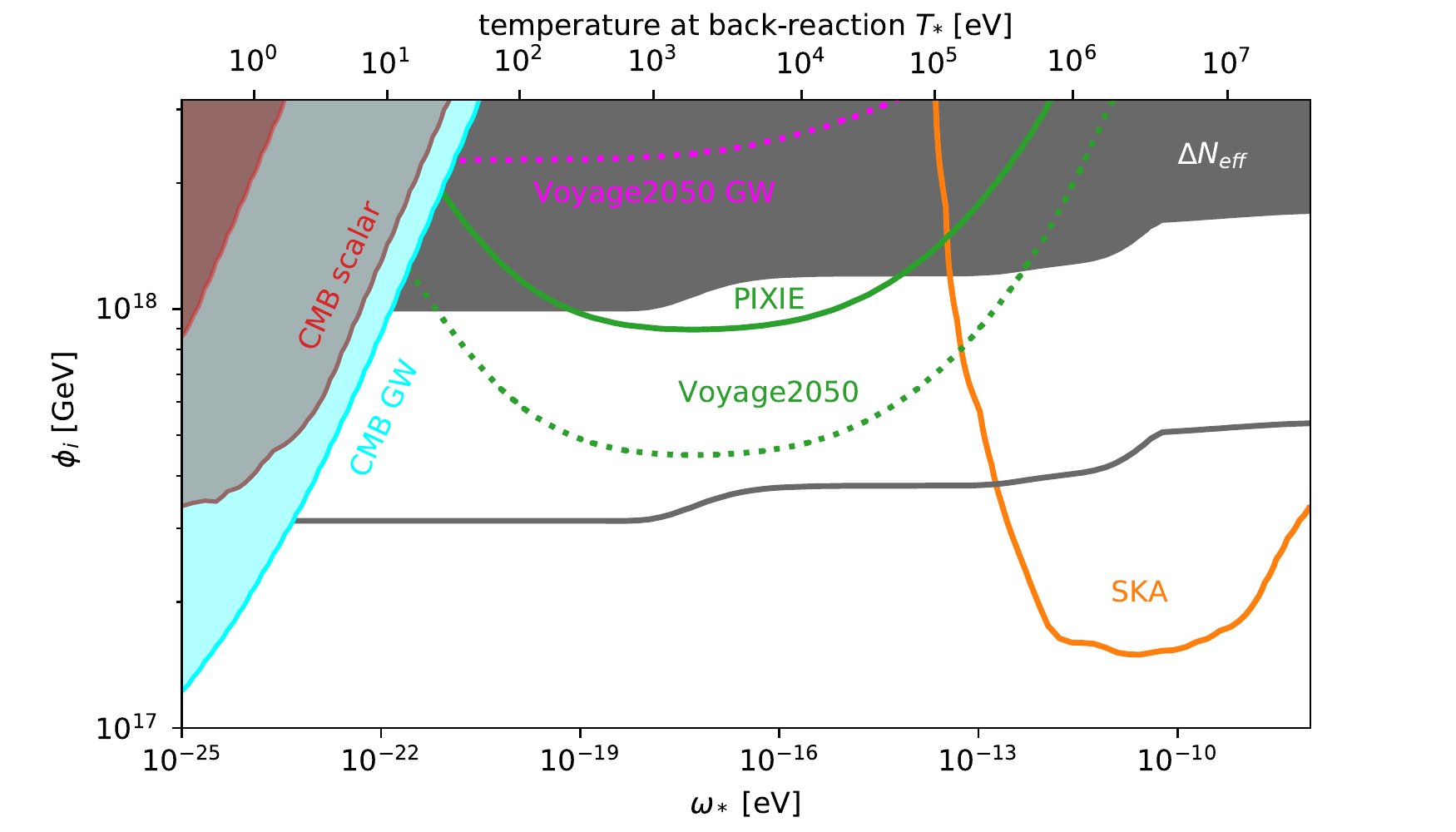}
        \caption{Constraints and future probes of the $\lambda\phi^4$-model introduced in \cref{sec:lambdaphi4}. Here $\omega_*$ determines the temperature $T_*$ at which the fluctuations come to dominate the energy in the dark sector and the initial amplitude $\phi_i$ gives the energy in the dark sector $\Omega_d$. 
        The gray area and line give the current and future bounds resulting from $\Omega_d$ increasing $\Neff$ at recombination.
        For $T_*$ below $\mathcal{O}(10^2\,\rm{eV})$ the scenario is constrained by fits to CMB fluctuations (red) as well as the non-observation of B-polarization modes (cyan). For the wide range of annihilation temperatures of $(10^2-10^7)\,$eV spectral distortions are able to probe this scenario with the future missions PIXIE (straight) and Voyage2050 (dotted). We show the bound including only the contribution from scalar acoustic waves in green and the one from only considering GWs in pink. At temperatures above $10^5\,$eV the model becomes testable by the future pulsar timing array SKA (orange line).}
        \label{fig:lphi4-parameterSpace}
    \end{figure}

To derive the bounds and reach of future experiments we use the spectra found in our lattice simulation and extrapolate them as $\propto k^3$ in the infrared and conservatively as 0 in the UV. The relic abundance, the energy in GWs and wave vectors $k$ are redshifted taking into account the changing number of relativistic degrees of freedom in the SM plasma in order to compare them to the future and present bounds mentioned above. The results are shown in \cref{fig:lphi4-parameterSpace}. We find that at low effective masses $\omega_*\lesssim10^{-22}$~eV the model is most stringently constrained by the non-observation of B-modes in the CMB (cyan). At intermediate values $10^{-23}~\text{eV}\lesssim\omega_*\lesssim10^{-13}$~eV the spectral distortions induced by acoustic waves will be detectable by future missions. We find that for this model, as for all other ones that we discuss, the contribution from the GWs to the distortion is negligible. At even larger masses, the model can be tested by SKA, but in the parameter space still allowed by the $\Neff$ constraints the signal is too weak to explain the recent findings of today's PTAs.

For this model, all bounds relying on scalar fluctuations are rather weak compared to the examples that we discuss below.
The reason is that the characteristic scale lies deep inside the horizon when the perturbations arise 
$k_*/(a_*H_*)=\mathcal{O}(100)$ and this factor enters with a power of -3 in the estimate \cref{eq:anaEst_stoch}. In the case of an axion coupled to a dark photon \cite{Kitajima:2017peg,Agrawal:2017eqm,Dror:2018pdh,Co:2018lka,Bastero-Gil:2018uel,Agrawal:2018vin,Machado:2018nqk,Machado:2019xuc,Ratzinger:2020koh,Madge:2021abk,Weiner:2020sxn,Adshead:2018doq,Cuissa:2018oiw,Kitajima:2020rpm,Co:2021rhi} featuring a similar instability this ratio is typically of the same order (possibly with the exception of \cite{Banerjee:2021oeu}) and we expect comparable results. The situation is different for the related scenario of axion fragmentation \cite{Fonseca:2019ypl,Jaeckel:2016qjp,Berges:2019dgr,Chatrchyan:2020pzh,Morgante:2021bks}, where this ratio can be of $\mathcal{O}(1)$ and we expect that the spectral distortions could be much larger than recently estimated in \cite{Eroncel:2022vjg}, where only the GWs were considered.

\subsection{Unstable Remnants of post-inflationary Symmetry Breaking }
\label{sec:domain_walls}
Symmetry breaking in the early universe is one of the most anticipated predictions for BSM physics, emerging in many extensions of the standard model e.g. \cite{Dine:1995uk,Kazanas:1980tx}. Here we will assume that the symmetry breaking takes place after inflation resulting in a universe filled with topological defects or a network of scaling seeds. We restrict ourselves to the study of domain walls from the breaking of a discrete symmetry, as well as cosmic strings resulting from the breaking of a global $U(1)$, which appear for instance, in axion-like particle (ALP) scenarios with post-inflationary Peccei-Quinn breaking.

\subsubsection{Domain Walls}
Domain walls (DWs)\cite{Vilenkin:1984ib,Vilenkin:1982ks,Kibble:1976sj} are two dimensional topological defects that emerge from the breaking of a discrete symmetry.
The parameter controlling the DWs dynamics after formation is the surface tension $\sigma$. By considering that every Hubble patch with volume $1/H^3$ contains a sheet of DW with area $1/H^2$ one can show that
\begin{align}
    \Omega_{DW}\simeq 0.5 \frac{\sigma}{\mPl^2 H}\,,
\end{align}
where the $\mathcal{O}(1)$ prefactor is inferred from simulations during radiation domination \cite{Saikawa:2017hiv}. This picture furthermore suggests that the system has $\mathcal{O}(1)$ density fluctuations at the horizon scale. As one can see the relative amount of energy in DWs grows as the universe cools down, leading to strict bounds on $\sigma$ in order to not over-close the universe. Observability therefore motivates a scenario in which the degeneracy of the vacua related by the symmetry is broken by an additional term in the potential $V_{\rm{bias}}$. The introduction of $V_{\rm{bias}}$ causes the walls to experience volume pressure, that leads to the annihilation of the network once the energy in the volume becomes comparable to the energy in the surface area of the DWs. 
In a radiation dominated universe the time of DW annihilation corresponds to the following temperature \cite{Saikawa:2017hiv} \begin{equation}
    T_{\rm{ann}} \approx 10 \hspace{1.0 mm} \rm{MeV} \left(\frac{\sigma}{ \rm{TeV^{3}}}\right)^{-\frac{1}{2}}\left(\frac{V_{bias}}{ \rm{MeV^{4}}}\right)^{\frac{1}{2}}.
\end{equation}

The GWs from annihilating DWs were first studied analytically \cite{Vilenkin:1981zs,Preskill:1991kd} and later on quantitatively using lattice simulations \cite{Hiramatsu:2013qaa,Hiramatsu:2010yz,Kawasaki:2011vv,Saikawa:2017hiv,Saikawa:2020duz}. 
On the lattice one finds that the GW spectrum is peaked at $k_{peak}=2\pi a_{\rm{ann}}H_{\rm{ann}}$ and the peak amplitude at emission is given as \cite{Hiramatsu:2013qaa}
\begin{align}
    \Omega_{GW,{\rm peak},{\rm ann}}\simeq 0.02\ \Omega_{DW,{\rm ann}}^2\,.
\end{align}
This is exactly what one expects from \cref{eq:roughEst_GW} for a source with dynamics on the horizon scale and $\mathcal{O}(1)$ density fluctuations. The shape of the spectrum is $\propto k^3$ for $k<k_{peak}$ and $\propto k^{-1}$ for $k>k_{peak}$. 

To estimate the acoustic energy induced by the DWs we use the spectrum given in \cref{eq:PowAnsatz} and set the normalisation to $A_{\delta_d}=1$. Since the spectrum peaks at $\approx2k_*$ we set $k_*=k_{peak}/2$. We have no reason to expect that the energy fluctuations $\delta_d(k,\tau)$ show a deterministic behavior and therefore use \cref{eq:anaEst_stoch}. The DWs are expect to source acoustic energy for the whole time that the network exists with the biggest contribution stemming from the time of annihilation when the relative energy in the network is largest. As a conservative estimate we only take this contribution into account and set $a_{*}H_{*}=a_{\rm{ann}}H_{\rm{ann}}$ and $\Omega_{d,*}=\Omega_{DW,{\rm ann}}$ in \cref{eq:anaEst_stoch}.

    \begin{figure}
        \centering
        \includegraphics[width=\textwidth]{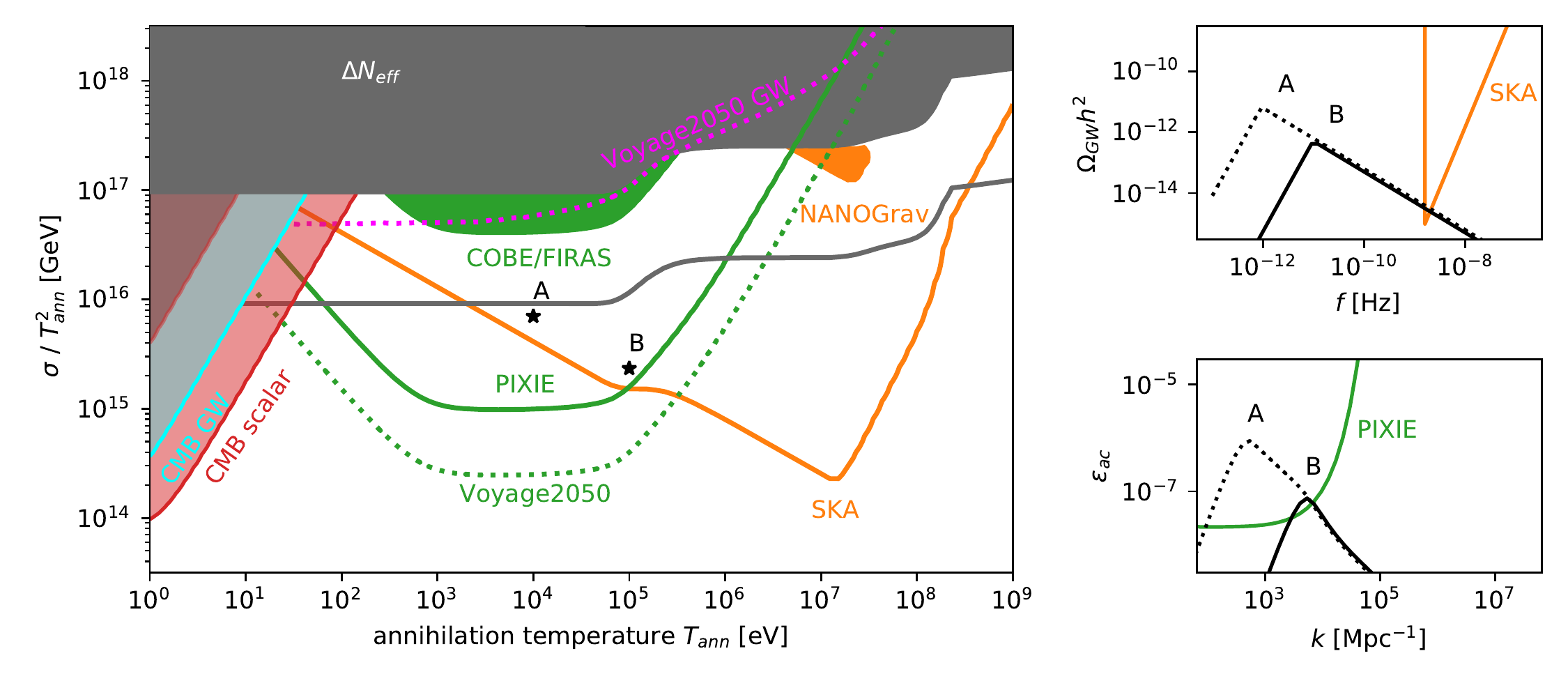}
        \vspace*{-2\baselineskip}
        \caption{Left: Constraints and future probes of domain walls in terms of the annihilation temperature of the network $T_{\rm{ann}}$ and the surface tension $\sigma$. See the text and \cref{fig:lphi4-parameterSpace} for discussion. We have picked two benchmarks A and B. Right: On the top we show the GW signal for the benchmarks in relation to the power law integrated noise of SKA. While SKA can not distinguish the two scenarios, the induced spectral distortion is drastically different as can be seen from the bottom panel.}
        \label{fig:DomainWalls}
    \end{figure}
    
In order to not over-close the Universe, we assume that the remnants of the DW network rapidly decay into dark, massless or light particles that consequently contribute to $\Neff$.
We take into account the appropriate redshift to arrive at the results shown in \cref{fig:DomainWalls}. We again find that the spectral distortions are dominantly produced through the damping of acoustic waves as can be seen by comparing the green and pink dotted line. The amount of acoustic energy and gravity waves is approximately the same $\epsilon_\text{ac}\sim\Omega_\text{GW}\propto \Omega_d^2$, but only a small fraction of the energy in gravity waves is injected into the photons, while all of the acoustic energy is converted when the modes cross the diffusion scale. 

Since the fluctuations are horizon sized, there is no suppression of the acoustic energy, resulting in stronger signals when compared to the model considered above, while still being compatible with the $\Neff$-bound. As can be seen from \cref{fig:DomainWalls}, in parts of the parameter space $T_{\rm{ann}}=10^3-10^5\,$eV the existing COBE/FIRAs data already constrains the distortions induced through acoustic waves (green area), slightly improving upon the bound from $\Neff$. 

On the right side of \cref{fig:DomainWalls} we sketch the GW spectrum for two benchmark points together with the expected sensitivity of SKA.\footnote{While for our parameter scan we use the exponential suppression in the UV from \cref{eq:PowAnsatz}, for this sketch we show a power law that we think is more realistic. The majority of the signal is due to the peak such that this introduces only a small uncertainty in the parameter scan.} Since PTAs are only sensitive to the UV tail of the spectrum, if the annihilation temperature is below $T_\text{ann}\lesssim10^7\,$eV, SKA is not able to distinguish between the two benchmarks. On the bottom we show the acoustic energy density spectra in relation to $\mu_\text{thr,PIXIE}/\mathcal{W}(k)$, where $\mu_\text{thr,PIXIE}=3\times 10^{-8}$ is the threshold for detection by PIXIE and $\mathcal{W}(k)$ is the window function given in \cref{eq:W_analytic}. Broadly speaking the overlap of the acoustic spectra with $\mu_\text{thr,PIXIE}/\mathcal{W}(k)$ gives the size of the signal compared to the threshold in accordance with \cref{eq:mu_from_W}. It becomes clear that the benchmarks, although indistinguishable by the SKA measurement, lead to drastically different $\mu$-distortions. This goes to demonstrate the role spectral distortions might have in the upcoming age of multi-messenger cosmology. The future missions PIXIE and Voyage2050 will go far beyond all other probes in this range.

\subsubsection{Global Strings}

Cosmic Strings (CSs) \cite{Kibble:1976sj,Vilenkin:1982ks}, one dimensional topological defects, are remnants of a spontaneous $U(1)$ symmetry breaking. The essential parameter controlling the dynamics of the strings is the symmetry breaking scale $f_{\phi}$ that determines the string tension. 
We will base our analysis in the following on the findings of \cite{Gorghetto:2021fsn}.
There it is found that the energy density in the string network is given as
 \begin{equation}\label{eq:string_energy}
     \Omega_{s}(a) \simeq1.0\cdot \log^2\left(\frac{f_{\phi}}{H(a)}\right) \left(\frac{ f_{\phi}}{\mPl}\right)^2,
 \end{equation} 
during radiation domination once the system has entered the scaling regime. The time-dependent logarithmic factor $\log(f_{\phi}/H)=\mathcal{O}(100)$ enters here, since parameters like the string tension and the number of strings per Hubble patch show this scaling in case of the breaking of a global $U(1)$ symmetry.
The existence and extent of this logarithmic dependence still remains debated for observables like the emitted GWs \cite{Gorghetto:2021fsn,Figueroa:2020lvo, Lopez-Eiguren:2017dmc,Chang:2021afa} though. According to \cite{Gorghetto:2021fsn} the energy of emitted GWs is given as
 \begin{equation}
     \Omega_{GW}(k) \simeq 0.2 \cdot \Omega_{s}^2|_{aH=k}\,.
 \end{equation}
Similar to the example of domain walls, we will again consider the possibility that an explicit breaking of the $U(1)$ symmetry enforces the annihilation of the network. This breaking is parameterized by the mass $m_{\phi}$ of the pseudo Nambu-Goldstone boson. Once Hubble drops to $H_*= m_{\phi}$, the field settles in its true minimum resulting in the formation of domain walls that collapse the network. Ref.~\cite{Gorghetto:2021fsn} finds that the GW spectrum features a peak at $k_\text{peak}=2\pi a_{*}H_{*}$ with the amplitude at the peak and higher frequencies given by the formula above and falls off as $k^3$ for lower frequencies. 

To determine the $\Neff$ bound we use that the energy in relativistic Nambu-Goldstone bosons at emission is \cite{Gorghetto:2021fsn}
\begin{align}
    \Omega_{\phi}\approx0.3\cdot \log^2\left(\frac{f_{\phi}}{H(a_*)}\right) \left(\frac{ f_{\phi}}{\mPl}\right)^2\,.
\end{align}
Part of these bosons will become non-relativistic and contribute to the DM density. We refer the reader to \cite{Gorghetto:2021fsn} for the derivation of the DM abundance as well as other bounds arising from structure formation (see also \cite{Murgia:2019duy,Rogers:2020ltq}).

\begin{figure}
    \def\sepf{0.7}
	\centering
    \includegraphics[scale=\sepf]{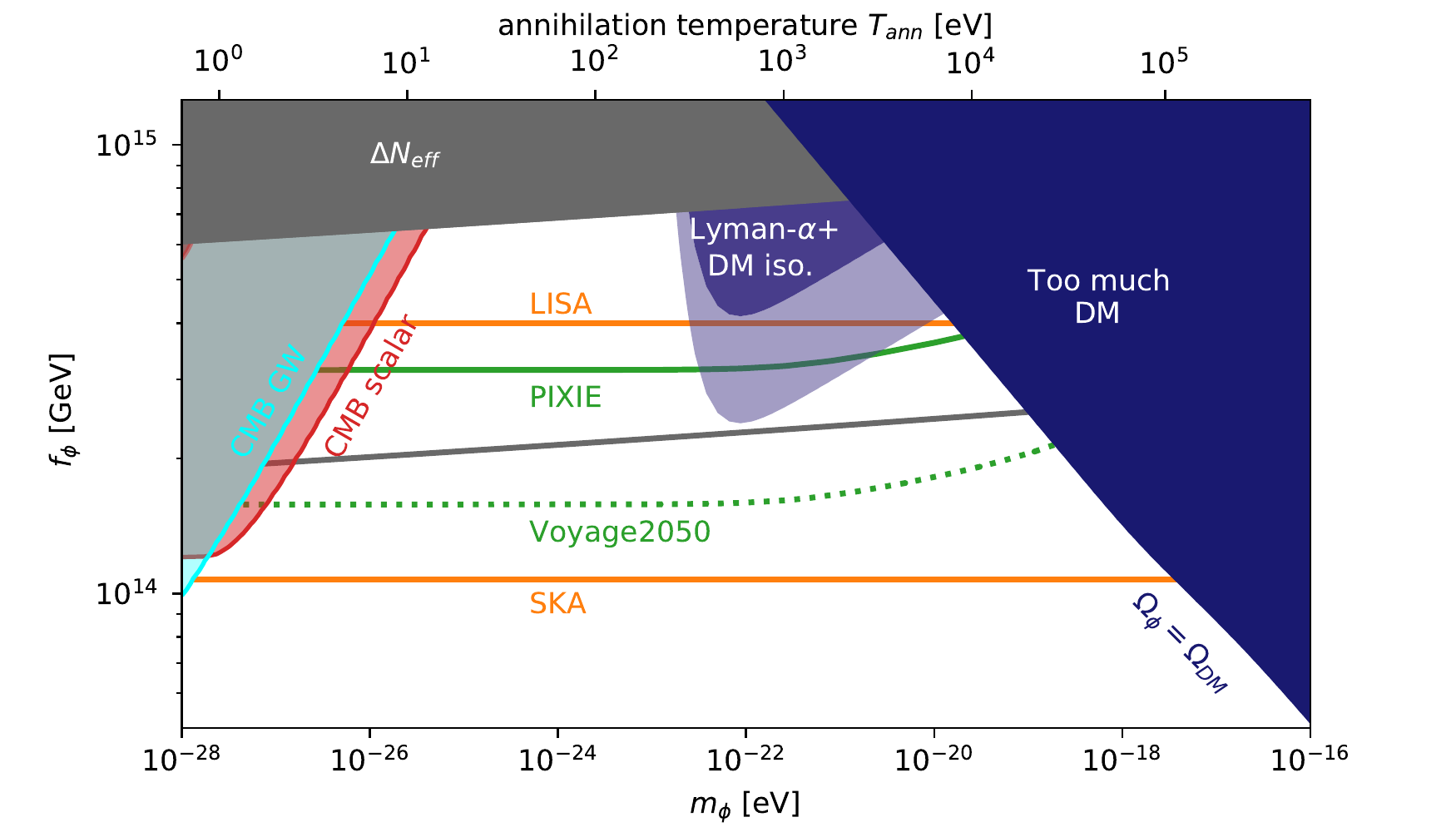} 
	\caption{Present and future constraints on cosmic string networks resulting from the breaking of a global $U(1)$ at the scale $f_{\phi}$. We assume that the network annihilates at a temperature $T_{ann}$ due to an explicit breaking of the symmetry parameterized by the mass of the resulting pseudo Nambu-Goldstone boson $m_\phi$. Additionally to the bounds shown in \cref{fig:lphi4-parameterSpace,fig:DomainWalls}, we also show the reach of the interferometer LISA searching for GWs. Further constraints on this parameter region arise due to the emitted axions making up a fraction of DM and featuring large isocurvature perturbations in conflict with Lyman-$\alpha$ observations (purple).
	For annihilation temperatures $T_{ann} \geq 10^{5} \rm{eV}$ the most severe constraint comes from overproducing axion DM. 
	Again  the great complementarity between GWs and spectral distortion experiments shall be emphasized, with the later being sensitive to the mass (at least for $m_\phi > 10^{-22}$~eV). }
	
	\label{fig:Strings}
\end{figure}

To estimate the acoustic energy induced by strings, we employ largely the same arguments and procedures as shown for the DWs: As a conservative estimate, we limit ourselves to the contribution of the strings leaving aside the bosons. We therefore plug $k_*=k_{peak}/2$, $A_{\delta_d}=1$ and \cref{eq:string_energy} into \cref{eq:anaEst_stoch} to get the estimate. To account for the continuous induction of acoustic energy during the scaling regime, we replace the exponential suppression for $k>k_{peak}$ by only a logarithmic dependence $\propto\log^4(k/k_{peak})$, in which we assumed that this is only due to the time dependence of $\Omega_{s}(a)$. The results are shown in \cref{fig:Strings}. We are additionally showing the reach of the future space-based interferometer LISA adopted from \cite{Gorghetto:2021fsn} as an orange line. The region in which the resulting DM overcloses the universe is shown in blue and the bounds arising from structure formation in purple. It should be mentioned that if our assumption of the density fluctuations being of horizon size at annihilation $k_*=\pi a_{*}H_{*}$ also holds true at late times, the stronger (purple shaded) bound is applicable. Interactions among the bosons could possibly relax this assumption though leading to a less stringent bound (solid purple). 

In the parameter space in which the pseudo Nambu-Goldstone bosons contribution to DM is not overclosing the universe, SKA and LISA only probe the UV tail of the GW spectrum, which renders them insensitive to the decay of the network and therefore $m_\phi$. Since for $m_\phi\gtrsim10^{-22}$eV the period in which acoustic waves are sourced is shortened, spectral distortions offer an opportunity to estimate or at least constrain $m_\phi$ in most of the parameter space with detectable GWs. 

Before moving on, let us compare our results to the ones obtained in \cite{Tashiro:2012pp,Amin:2014ada} for non-decaying networks of scaling seeds. Both papers consider the spectral distortions due to gravitationally induced acoustic waves in the baryon-photon fluid, just as we do in this paper. Ref.~\cite{Tashiro:2012pp} considers the case of cosmic strings and found that only $\mu\approx 10^{-13}$ can be reached without being in tension with CMB observations, while our analysis suggests that distortions as large as $\mu\approx10^{-9}-10^{-10}$ are possible.\footnote{It should be mentioned that \cite{Tashiro:2012pp} concerns the case of local strings. Global strings effectively radiate off their energy as Goldstone bosons preventing the formation of small loops, while local strings predominantly decay into gravitational waves and therefore form small loops. This makes the two scenarios drastically different when it comes to the emission of gravitational waves. We seem to agree with \cite{Tashiro:2012pp} that the gravitational drag from a surrounding fluid only concerns the largest scales of the respective network, as can be seen from our estimate \cref{eq:anaEst_stoch} decaying as $k^{-3}$ and the integration over the loop-length in \cite{Tashiro:2012pp} being dominated by the largest loops. Therefore, in this regard the two scenarios should be similar and this does not necessarily explain the tension between our results. Ref. \cite{Tashiro:2012pp} introduces a wiggliness parameter that potentially enhances the distortion. If this parameter is proportional to $\log(f_{\phi}/H)$ the two estimates can be reconciled.}
In \cite{Amin:2014ada} the breaking of an $\mathcal{O}(N)$ symmetry with $N\geq4$ was studied that features no topological artefacts but a network of scaling seeds with quasi constant $\Omega_d$ and dynamics of horizon size. This system is therefore very similar to cosmic strings. They find that present CMB bounds allow for $\mu\approx10^{-9}$ for non-decaying networks in good agreement with our result.

\subsection{Phase transitions}
\label{sec:phase_transitions}

\begin{figure}
    \def\sepf{0.7}
	\centering
    \includegraphics[scale=\sepf]{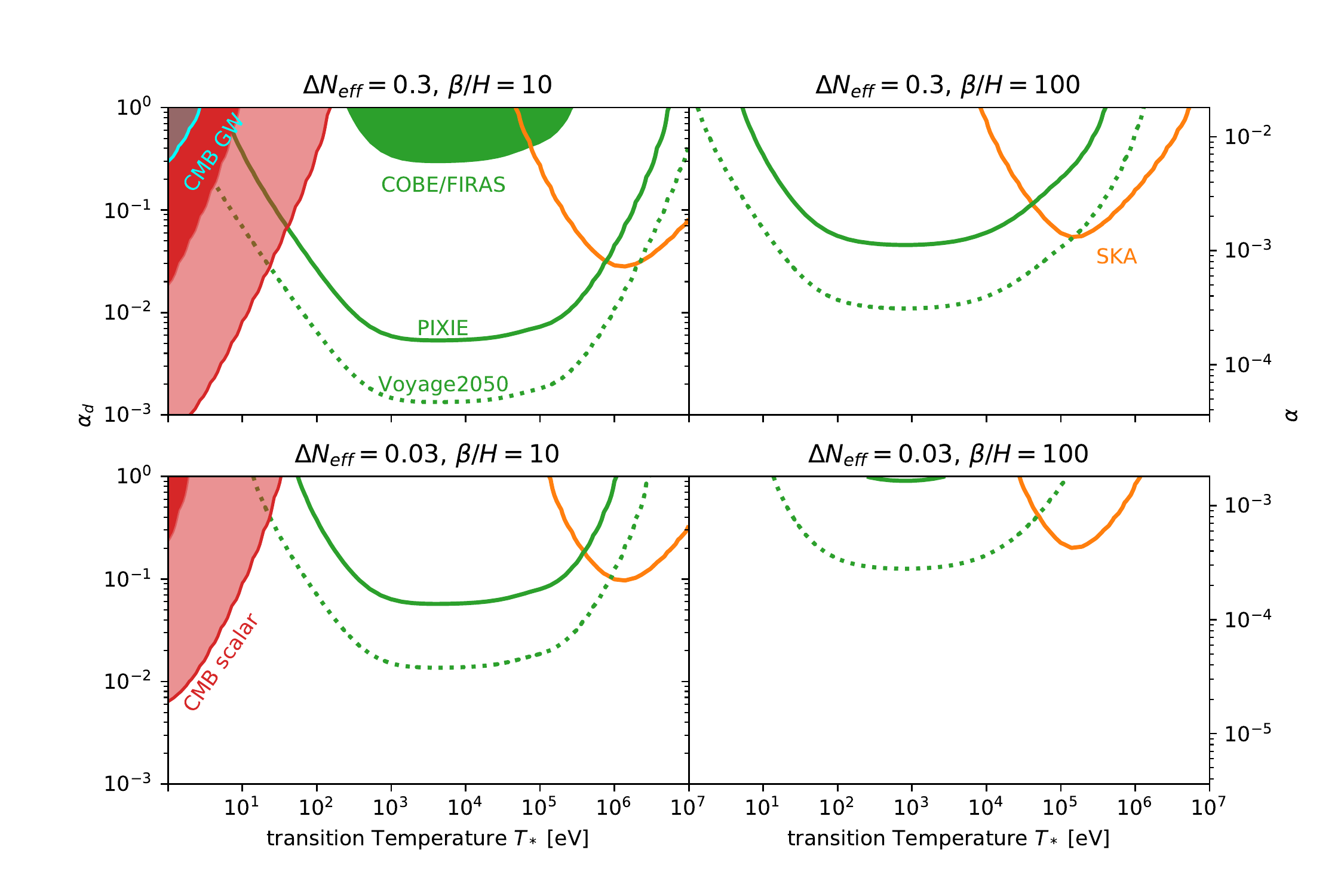} 
    \vspace*{-2\baselineskip}
	\caption{ Current and future constraints on a phase transition in a dark sector, in terms of the SM plasma temperature at the time of the transition and the released energy relative to the energy in the dark fluid $\alpha_{d}$, for different values of  $\Delta N_{eff}$ and the inverse duration of the phase transition $\frac{\beta}{H}$. We only take into account the effects of the sound waves in the dark fluid caused by the transition. At temperatures $T_{*}= 1-10\,$eV the scenario is constrained by CMB fluctuations in (red) and the non-observation of B-mode polarization in the CMB in (cyan). In the temperature range of $T_{*}= 10-10^{6}\,$eV  spectral distortions from acoustic waves (green) can probe the scenario. Strong phase transitions in a dark sector saturating the $\Neff$ bound  can already be constrained by the COBE/FIRAS results. At temperatures $T_{*}= 10^5-10^7\,$eV the scenario can in the future be detected by SKA, while the present $\Neff$ bound (top row) rules out the NANOGrav signal being generated this way. For convenience we have converted $\alpha_d$ to $\alpha=\Omega_d\alpha_d/(1+\alpha_d)$ on the left y-axis.}
	\label{fig:Phasetransi}
\end{figure} 

Many theories of BSM physics predict a first-order phase transition (FOPT) in the post-inflationary universe (see e.g.~\cite{Witten:1984rs,Hogan:1983ixn,Kuzmin:1985mm}). These FOPT proceed through bubble nucleation and bubble collisions at relativistic speeds and are therefore a great source of GWs \cite{Hogan:1986qda,Kamionkowski:1993fg}. We consider the case in which the FOPT takes place in a purely gravitationally coupled sector as in \cite{Schwaller:2015tja,Breitbach:2018ddu,Fairbairn:2019xog} and only consider the sound wave contribution to the GW spectrum \cite{Caprini:2015zlo,Caprini:2019egz}. While this is often the dominant source of GWs from thermal FOPTs, this is a conservative estimate of the GW spectrum since it neglects contributions from bubble collisions and turbulence.

The parameters describing such a system are the energy density in the dark sector $\Omega_d$ that can readily be exchanged for the contribution to $\Neff$ using \cref{eq:Neff}, the amount of energy freed in the phase transition relative to the one in the dark fluid $\alpha_d$,
\footnote{The situation commonly discussed only concerns the case where the universe is filled with one fluid. When dealing with multiple fluids the introduction of $\alpha_d$ and $\Omega_d$ is necessary as opposed to only using $\alpha= \Omega_{d}\alpha_{d}/(1+\alpha_{d})$. Since the bubble walls only couple to the fluid in the dark sector, the matching conditions are only imposed on the dark sector and as a result the efficiency of generating sound waves depends on $\alpha_d$ as opposed to $\alpha$ \cite{Fairbairn:2019xog}.}
the time of the transition which we will give as the temperature $T_*$ of the SM plasma at the time,\footnote{The dark sector, if it is thermal, must not have the same temperature as the SM plasma. If it posses one relativistic degree of freedom its temperature is necessarily smaller.  See \cite{Fairbairn:2019xog,Breitbach:2018ddu} for further details.} as well as the inverse time scale of the transition $\beta$. To keep our discussion simple we further set the wall velocity $v_w\simeq1$ and restrict us to a speed of sound  $c_{d}=1/\sqrt{3}$ in the dark sector,
\footnote{In figure \ref{fig:ac_energy_suppression} we show the effect of varying the sound velocity and find that the suppression in the amount of acoustic energy is small as long as the changes don't exceed $\approx k_{*}/(a_*H_*)\approx\beta/H_*$. It has been found though that even small changes in both the sound velocity and the wall velocity can have an significant impact on the efficiency factor $\kappa$, entering both the GW and acoustic energy estimate.\cite{Espinosa:2010hh}}
which allows us to estimate its acoustic energy relative to its total energy as
\begin{align}
    \epsilon_{\text{ac},d}=\frac{\rho_{\text{ac},d}}{\rho_{d}}=\frac{\kappa(\alpha_d) \alpha_d}{1 + \alpha_d}\,;\qquad \kappa(\alpha_d)=\frac{\alpha_d}{0.73+0.083\sqrt{\alpha_d}+\alpha_d}\,,
\end{align}
where $\kappa$ gives the efficiency factor for converting the released energy into sound waves as found in \cite{Espinosa:2010hh}.\\\\
The energy density of GWs coming from the sound waves which are emitted from a dark sector with nucleated bubbles of sub-horizon size is \cite{Hindmarsh:2015qta}
\begin{equation}\label{eq:GWPT}    
\Omega_{GW}(k) \simeq 0.16\left(\frac{k}{k_\text{peak}}\right)^3\left(\frac{7}{4+3(k/k_\text{peak})}\right)^{7/2}\cdot \Omega_{d_{*}}^{2}\cdot\frac{H_*}{\beta} \cdot\left(\frac{\kappa(\alpha_d) \alpha_d}{1 + \alpha_d }\right)^{2}\,.
\end{equation}
The first term is again an $\mathcal{O}(1)$ prefactor for $k=k_\text{peak}=2a_{*}\beta/\sqrt{3}$ and determines the shape of the spectrum, while we can identify the other terms with the factors in the rough estimate of \cref{eq:roughEst_GW}.

The density fluctuations in the relativistic dark sector, just as in the baryon-photon fluid, are related to the acoustic energy via $A_{\delta_d}=\langle\delta_d^2\rangle=4\epsilon_{\text{ac},d}$ in the virial limit. Again we set $k_*=k_\text{peak}/2$, but in this case we use \cref{eq:anaEst_det} to determine the gravitationally induced acoustic waves. This is justified, since one expects that $\delta_d$ only shows a stochastic behavior for a time $\approx 1/\beta$ while the walls are present and proceeds with the deterministic propagation of sound waves for the remaining Hubble time following the transition.

The ratio $\beta/H_*$ determines if the PT completes mainly driven by the expansion of a few nucleated bubbles or by the nucleation of new bubbles everywhere in space. Large values of $\beta/H_*$ correspond to faster nucleation rates which means that more bubbles will nucleate inside the Hubble horizon until the PT has completed, and hence their bubble radii get smaller. One expects an inverse relation between $\beta/H_*$ and the amplitude of the GW spectrum. Since $\beta/H_*$ determines the average bubble radius at the time of collision, it also controls the peak frequency of the GW spectrum. For very strong FOPTs one might have to reformulate the definition of $\beta/H_*$ as it may become inappropriate and was emphasized in \cite{Huber:2007vva,Jinno:2017ixd}.

In Fig \ref{fig:Phasetransi} we show our results. Similar to the previous examples we find that spectral distortions bridge the gap between phase transitions detectable by CMB fluctuations ($T_*=1-10\,$eV) and by PTAs such as SKA ($T_{*}= 10^5-10^7\,$eV). The probes relying on scalar mediation are however particularly strong for $\alpha_d\ll1$. In this case the fluctuations in the dark sector are small $\mathcal{P}_{\delta_d}\ll1$, which suppresses the GWs relative to the sourced acoustic waves, as one can see by comparing \cref{eq:roughEst_GW} with \cref{eq:anaEst_stoch,eq:anaEst_det}.

\subsection{Comment on Directly Coupled Sectors}
\begin{figure}
    \def\sepf{0.65}
	\centering
    \includegraphics[scale=\sepf]{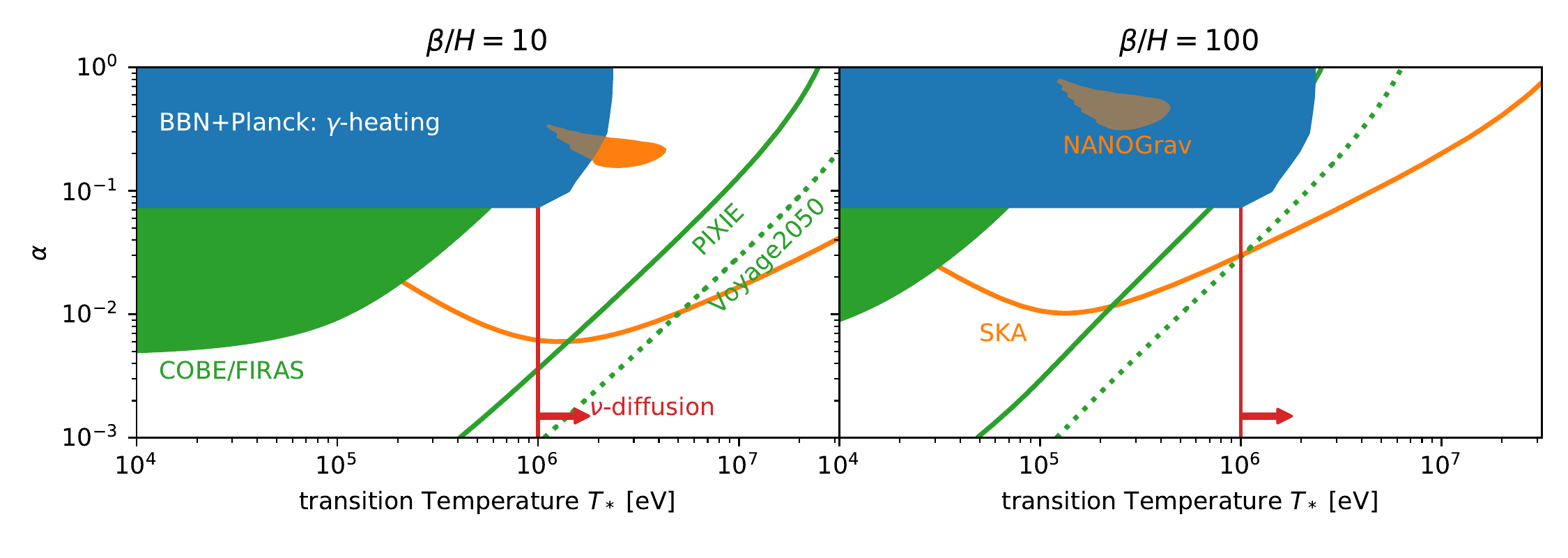}        \vspace*{-\baselineskip}
	\caption{Current and future constraints on a first order phase transition at temperature $T_{*}$ releasing a relative energy $\alpha$ into the SM-plasma. For temperatures below $\approx \SI{2}{MeV}$ the released energy $\alpha$ leads to tensions in BBN and CMB measurements of the baryon to photon ratio (blue). The sound waves caused by the phase transition source GWs that can explain the NANOGrav hint (orange, filled) and in the future can be detected over a wide range of parameter space (orange line). The green area and lines show the current and future sensitivity to spectral distortions caused by the sound waves. At temperatures above $\approx \SI{1}{MeV}$ the sound waves and therefore spectral distortions are expected to be reduced due to damping by neutrino diffusion.}
	\label{fig:Phasetransi_visible}
\end{figure} 

Clearly, one can obtain stronger bounds from spectral distortions if there are additional interactions between the new physics sector and the SM plasma besides gravity. In this sense, the bounds presented above can be interpreted as lower limits since gravity is always present. Studies of such scenarios will need to  make specific assumptions about the nature of the coupling and can therefore not be carried out in a model-independent way as we did in \cref{sec:source_mu} and \cref{sec:sound_energy}, but case by case as was done e.g.~for cosmic strings~\cite{Tashiro:2012nb, Anthonisen:2015tra}. 

In light of the recent findings by PTAs, we here want to briefly comment on the case in which a phase transition directly causes sound waves in the SM plasma and subsequently emits GWs. From the previous section, we saw that GWs from PTs in decoupled sectors can not explain the NANOGrav data due to the $\Neff$ bound. If the energy released in the PT is instead directly deposited into SM degrees of freedom, the $\Neff$ constraint does not apply. One can therefore consider the case where the PT sources sound waves in the SM plasma directly, which then produce the GWs.

Since the walls now directly source the acoustic energy in the baryon-photon fluid, we no longer rely on the gravitational coupling. Instead the acoustic energy spectrum can be constructed from the total acoustic energy given in \cite{Espinosa:2010hh} and assuming a white noise spectrum
\begin{align}
    \epsilon_\text{ac}(k)=\frac{\kappa(\alpha)\alpha}{1+\alpha}\sqrt{\frac{2}{\pi}}\frac{k^3}{k_*^3}\exp\left(-\frac{k^2}{2k_*^2}\right)\,,
\end{align}
with $k_*=a_{*}\beta/\sqrt{3}$. To estimate the GW signal strength, we use the formulas from the previous section, setting $\Omega_d\rightarrow1$ and replacing $\alpha_d\rightarrow\alpha$. With these changes we arrive back at the expression for the case where there is only one fluid present \cite{Hindmarsh:2015qta}.\footnote{For transition temperatures below $\approx 1$~MeV the neutrinos are decoupled and there are technically two sectors. Since the energy in the baryon-photon fluid is still $\Omega_\gamma\approx1$, we make this simplifying assumption.} 

The results are shown in \cref{fig:Phasetransi_visible}. An energy injection around or after BBN at $T\approx1$~MeV leads to a possible tension between the baryon to photon ratio obtained from BBN and CMB measurements. The resulting current bound on $\alpha$ and its temperature dependence has been investigated in~\cite{Bai:2021ibt}, and we show it in blue. As can be seen, this bound already excludes a decent chunk of the $2\sigma$-region of the NANOGrav fit (orange, filled). However we find that the remaining region can be probed by future distortion experiments, provided that our above estimate holds. Furthermore we obtain a significant overlap of the parameter space testable by SKA and spectral distortions.

The previously mentioned conclusions come, however, with the following caveat: At the beginning of BBN around $T\approx1$~MeV, the neutrinos decoupled from the rest of the SM plasma. Similar to the decoupling of photons, one has to expect that all perturbations on sub-horizon scales might be significantly damped due to the diffusion of neutrinos. We anticipate that this effect would reduce the reach of distortion searches for PT temperatures above $1$~MeV. The previously mentioned effect covers a significant region of the viable parameter space shown by the red line in \cref{fig:Phasetransi_visible}. For transition temperatures close to $1$~MeV, in that region, a reduction of the GW amplitude is plausible as well as their emission and the damping by $\nu$-diffusion are taking place simultaneously. We leave a detailed study of these effects to future work.
\section{Conclusion}
Spectral distortions caused by gravitationally induced acoustic waves in the baryon-photon fluid provide a powerful probe of new physics. We showed that the induction of the acoustic waves via the gravitational coupling and the subsequent damping due to diffusion can be separated. Since the latter is completely determined by SM physics, the $\mu$-distortions can be calculated solely from the amplitude or equivalently the energy of the waves (\cref{eq:mu_from_ac_power}). 

We here for the first time presented a general, analytic estimate of this acoustic energy caused by a purely gravitationally coupled sector (\cref{eq:anaEst_stoch,eq:anaEst_det}). This estimate only relies on a few parameters describing the dynamics of the dark sector. The most important amongst them are the amount of energy constituted by the dark sector, the amplitude of its energy fluctuations as well as the ratio between the typical length scale of the fluctuations and the horizon when the fluctuations are generated. While we restricted ourselves to studying sectors with relativistic dynamics, we expect that the results from \cref{sec:sound_energy} can be generalized to non-relativistic dynamics easily.

We continued by studying a particularly easy model consisting of two scalars in all detail. Solving for the dynamics in the dark sector using a lattice simulation and afterwards calculating the acoustic waves numerically allowed us to get the acoustic energy from first principles. We compared the result obtained in this way to our analytic estimate and found agreement to within $\mathcal{O}(1)$ factors, in the peak region even to within 20\%. 

Finally our analytic estimate allowed us to evaluate the spectral distortion signal for several dark sector scenarios.
We were able to demonstrate that dark sectors with energy fluctuations present at temperatures between $\approx \SI{10}-\SI{e6}{eV}$ are either already constrained by the non-observation of spectral distortions or can be probed in the future.
These regions of parameter space are not constrained by other probes relying only on gravitational coupling like $\Neff$, CMB fluctuations and GWs.
A particular interesting opportunity lies in the possible interplay with GW observations by PTAs. Here spectral distortions might be key in lifting observational degeneracies on the parameter space of various models.


\section*{Acknowledgements}
We would like to thank Daniel Figueroa for practical help with the simulations as well as Valerie Domcke, Edward Hardy and Peera Simakachorn for useful discussions.
Additionally, the authors gratefully acknowledge the computing time granted on the Mogon supercomputer at Johannes Gutenberg University Mainz (hpc.uni-mainz.de). Work in Mainz is supported by the Deutsche Forschungsgemeinschaft (DFG), Project No. 438947057 and by the Cluster of Excellence “Precision Physics, Fundamental Interactions, and Structure of Matter” (PRISMA+ EXC 2118/1) funded by the German Research Foundation (DFG) within the German Excellence Strategy (Project No. 39083149).

\appendix

\section{Numerics of Fluctuation Dynamics}
\label{sec:TCA_nuDyn}
To find the $\mu$-distortion and/or acoustic energy $\epsilon_\text{ac}^{\lim{}}$ we solve  \cref{eq:phi,eq:psi,eq:delta_gamma,eq:v_gamma} numerically, supplemented with the free streaming neutrinos. To describe the neutrinos we have to go past the first three moments of the phase-space distribution $\delta,v$ and $\sigma$. We adopt the conventions used in  \cite{Dodelson:2003ft} and write density fluctuations as deviations in temperature $\Theta(\vect{x},\vect{\hat{p}},\tau) = \frac{\delta T}{\Bar{T}}(\vect{x},\vect{\hat{p}},\tau)$. After Fourier transforming $\vect{x}\rightarrow\vect{k}$ the momenta of the distribution are defined as
\begin{align}
 \theta_l=i^l\int^1_{-1}\frac{d\eta}{2}P_l(\eta)\theta(\eta)\,,
\end{align}
where $P_l$ is the $l$th Legendre polynomial and $\eta=\vect{\hat k}\cdot\vect{\hat p}$. The first three moments  can be related via $\delta=4\theta_0$, $v=3\theta_1$ and $\sigma=2\theta_2$ to the definitions used in the main text. For free streaming neutrinos the dynamics of $\Theta_n(\vect{x},\vect{\hat{p}},\tau)$ are described by the Boltzmann equation without a scattering term, which in the expansion introduced above becomes \cite{Dodelson:2003ft,Ma:1995ey}
\begin{align}
 \dthetan{0}+k\thetan{1}&=-\dot\phi\label{eq:neutrino_evolution}\\
 \dthetan{1}-k\left(\frac13 \thetan{0}-\frac23 \thetan{2}\right)&=\frac{k}{3}\psi\\
 \dthetan{l}-k\left(\frac{l}{2l+1} \thetan{l-1}-\frac{l+1}{2l+1} \thetan{l+1}\right)&=0\,;\qquad\ l\geq2\,.
\end{align}
We truncate this hierarchy by neglecting moments $l>l_\text{max}=4$ and follow \cite{Ma:1995ey} to close the system of equations using
\begin{align}
    \thetan{l_\text{max}+1}=\frac{2l_\text{max}+1}{k\tau}\thetan{l_\text{max}{}}-\thetan{l_\text{max}-1}\,.
\end{align}

As initial conditions we consider the gravitational potentials as well as all the fluctuations in the SM sector to be zero and supply the fluctuations of the dark sector either as an analytic Ansatz or as an interpolation of the values we get from the lattice simulation. To calculate $\epsilon_\text{ac}^{\lim{}}$ we set $\dtc^{-1}=0$ to decouple the generation of the acoustic energy completely from the damping. When calculating the $\mu$-distortion directly without the separation approximation derived in \cref{sec:source_mu}, one needs to include $\dtc$ when solving the differential equations and calculate the time integral in \cref{eq:mu_from_ac_power} numerically using the solutions.

\section{Free Scalar Field}
\label{sec:FreeScalarField}
Below we calculate the energy fluctuations and their autocorrelation functions for a single relativistic scalar field $\phi$ with Gaussian fluctuations
\begin{align}
    \Pow_{\phi}(k)=A_{\phi}\sqrt{\frac{2}{\pi}}\frac{k^3}{\tilde{k}_*^3}\exp\left(-\frac{k^2}{2\tilde{k}_*^2}\right),
\end{align}
where $1/\tilde{k}_*$ is the characteristic length scale of fluctuations in the field as opposed to the energy density. We assume that the modes of the field are virialized such that $\Pow_{\dot\phi}(k)=\omega_k^2\Pow_{\phi}(k)$, where $\omega_k^2=k^2+m^2$ is the frequency of the respective mode. For a free scalar field the mode functions follow the equation of motion of an unperturbed harmonic oscillator, which is why the autocorrelation of both $\phi$ and $\dot\phi$ is given as $\cos(\omega_k t)$, while the cross-correlation is given as
\begin{align}
    \langle \phi(\vect{k},t)\dot\phi^*(\vect{k}',t') \rangle=\frac{2\pi^2}{k^3}\Pow_{\phi}(k)\, \omega_{k}\sin\left(\omega_k(t-t')\right)\, (2\pi)^3\delta^{(3)}(\vect{k}-\vect{k}')\,.
\end{align}

\subsection*{Energy Fluctuations}
The energy density of the field is
\begin{align}
    \rho_{\phi}=\frac{1}{2}\left(\dot\phi^2+\vect{\nabla}\phi^2+m^2\phi^2\right)
\end{align}
and its Fourier coefficients are given as
\begin{align}
    \rho_{\phi}(\vect{k})=\frac{1}{2}\int\frac{d^3p}{(2\pi)^3}\, \dot\phi(\vect{p})\dot\phi(\vect{k}-\vect{p})+ \left[m^2-\vect{p}\cdot(\vect{k}-\vect{p})\right]\phi(\vect{p})\phi(\vect{k}-\vect{p})\,.
\end{align}
The mean energy density can be calculated as
\begin{align}
    \overline{\rho}_{\phi}=\frac{1}{V}\langle\rho_{\phi}(\vect{k}=0)\rangle=
    \begin{cases}
        3A_{\phi}\tilde{k}_*^2 &\omega_k\approx k\\
        A_{\phi}m^2 &\omega_k\approx m
    \end{cases}
\end{align}
for the relativistic and non-relativistic case, respectively, and $V$ denotes the volume one is averaging over. When calculating the correlation of energy fluctuations $\langle\rho_{\phi}(\vect{k})\rho^*_{\phi}(\vect{k}')\rangle$ we encounter the following kinds of correlators between Gaussian variables:
\begin{align}
    &\langle\phi(\vect{p})\phi(\vect{k}-\vect{p})\phi^*(\vect{p}')\phi^*(\vect{k}'-\vect{p}')\rangle\\
    =& \langle\phi(\vect{p})\phi^*(\vect{p}')\rangle\langle\phi(\vect{k}-\vect{p})\phi^*(\vect{k}'-\vect{p}')\rangle+\langle\phi(\vect{p})\phi^*(\vect{k}'-\vect{p}')\rangle\langle\phi(\vect{k}-\vect{p})\phi^*(\vect{p}')\rangle\\
    =& (2\pi)^3\delta^3(\vect{k}-\vect{k}')\left[(2\pi)^3\delta^3(\vect{p}-\vect{p}')+(2\pi)^3\delta^3(\vect{p}-(\vect{k}'-\vect{p}'))\right]\frac{2\pi^2}{p^3}\Pow_{\phi}(\vect{p})\frac{2\pi^2}{|\vect{k}-\vect{p}|^3}\Pow_{\phi}(\vect{k}-\vect{p})\,,
\end{align}
where we assumed $\vect{k}\neq0$ and therefore $\langle\phi(\vect{p})\phi(\vect{k}-\vect{p})\rangle=0$. Putting it all together we arrive at
\begin{align}
    \Pow_{\rho_\phi}(k,t,t+\Delta t)&=\frac{k^3}{2\pi^2} \frac{1}{2}\int\frac{d^3p}{(2\pi)^3} \frac{2\pi^2}{p^{3}}\Pow_{\phi}(p) \frac{2\pi^2}{|\vect{k}-\vect{p}|^3}\Pow_{\phi}(k-p)\cdot\label{eq:UTCrhoFree}\\ &\hspace{2cm}\bigg[\left(\omega_p^2\omega_{k-p}^2+(m^2-\vect{p}\cdot(\vect{k}-\vect{p}))^2\right) \cos(\omega_p\Delta t)\cos(\omega_{k-p}\Delta t)+\nonumber\\
    &\hspace{2.2cm}2\omega_p\omega_{k-p}(m^2-\vect{p}\cdot(\vect{k}-\vect{p})) \sin(\omega_p\Delta t)\sin(\omega_{k-p}\Delta t)\bigg]. \nonumber
\end{align}
We can further evaluate this expression for $k\ll\tilde{k}^*$. In this case we approximate $\vect{p}=\vect{p}-\vect{k}$ except when evaluating $\omega_p$ and $\omega_{k-p}$ in the sine and cosine, since we want to keep track of the time evolution. The $p$-integral is dominated by modes with $p\approx\tilde{k}^*$ and we therefore approximate
\begin{align}
    \Delta\omega=\omega_{p}-\omega_{k-p}\approx
    \begin{cases}
        \vect{k}\cdot\vect{p}/|\vect{p}| &\omega_{\tilde{k}_*}\approx \tilde{k}_*\\
        \vect{k}\cdot\vect{p}/m &\omega_{\tilde{k}_*}\approx m\,.
    \end{cases}
\end{align}
We then find by using trigonometric identities
\begin{align}
    \Pow_{\rho_\phi}(k,t,t+\Delta t)&\approx\frac{k^3}{2\pi^2} \int^{\infty}_{0} d\log p\ \frac{2\pi^2}{p^3} \Pow^2_{\phi}(p)\, \omega_p^4 \int_{S^2}\frac{d\Omega_p}{4\pi} \cos(\Delta\omega\Delta t)\,.\label{eq:UTCrhoFreeIR}
\end{align}
The last integral in this expression is the autocorrelation function of the energy fluctuations. In the relativistic case it does not depend on $|\vect{p}|$, while in the non-relativistic we can approximate $|\vect{p}|\approx \tilde{k}_*$ and introduce the typical velocity of energy transport in the dark sector as $c_d=\tilde{k}_*/m$ to find
\begin{align}
    \mathcal{A}_{\delta_\phi}(k,\Delta t)=
    \begin{cases}
        \text{sinc}(kt) &\omega_{\tilde{k}_*}\approx \tilde{k}_*\\
        \text{sinc}(c_d kt)  &\omega_{\tilde{k}_*}\approx m\,.
    \end{cases}
\end{align}
We argued in the main text that the only relevant time scale for the autocorrelation of the energy density is $c_d/k$, with $c_d$ the typical velocity of energy transport. Here we showed this explicitly.

\subsection*{Shear}
The space-space part of the energy momentum tensor of a scalar field is given by
\begin{align}
    T_{ij}\approx\vect{\nabla}_i\phi\vect{\nabla}_j\phi\,,\label{eq:shear_scalarField}
\end{align}
where we neglected contributions proportional to $g_{ij}$ that exclusively contribute to the trace. We find the shear by going to Fourier space and projecting out the longitudinal traceless component
\begin{align}
    \sigma_{\phi}(\vect{k})=\frac{1}{\overline\rho_\phi+\overline p_\phi}\left[\frac{1}{k^2}k_iT_{ij}(\vect{k})k_j-\frac{1}{3}T_{ii}(\vect{k})\right].
\end{align}
From there the steps are the same as for the energy density and we arrive at
\begin{align}
    \Pow_{\sigma_\phi}(k,t,t+\Delta t)&=\frac{1}{(\overline\rho_\phi+\overline p_\phi)^2}\frac{k^3}{2\pi^2} 2\int\frac{d^3p}{(2\pi)^3} \frac{2\pi^2}{p^{3}}\Pow_{\phi}(p) \frac{2\pi^2}{|\vect{k}-\vect{p}|^3}\Pow_{\phi}(k-p)\cdot\label{eq:UTCsigmaFree}\\ 
    &\hspace{2cm}\bigg[\left(\hat{\vect{k}}\vect{p}\right)\left(\hat{\vect{k}}(\vect{p}-\vect{k})\right)-\frac{1}{3} \vect{p}(\vect{p}-\vect{k})\bigg]^2\cos(\omega_p\Delta t)\cos(\omega_{k-p}\Delta t)\,. \nonumber
\end{align}
When we expand the cosines again for $k\ll\tilde{k}_*$, we find
\begin{align}
    \cos(\omega_p\Delta t)\cos(\omega_{k-p}\Delta t)=\frac{1}{2}\left(\cos(2\omega_{\tilde{k}_*}t)+\cos(\Delta\omega t)\right)\,.
\end{align}
To arrive at the autocorrelation function one would need to carry out the integration. But we are content here with only showing that for a non-conserved quantity like the shear indeed both time scales $1/\tilde{k}_*$ and $1/k$ enter. This can already be seen from the above expression with the $\cos(2\omega_{\tilde{k}_*}t)$ term not canceling.

\section{Simulation of the $\lambda\phi^4$-Theory}
\label{sec:NumericsPhi4}
\subsection*{Energy Density}
When working with a Velocity Verlet (VV) or Runge Kuta type integrator both the fields and their time derivatives or equivalently momenta are given at the same point in time. This is not the case for a leapfrog integration scheme, but can easily be achieved by introducing a half time step for the momenta, to recover the Velocity Verlet procedure. From these one can calculate the total energy as discussed in e.g. \cite{Figueroa:2020rrl}. In principle a generalization to an energy density is straight forward. The only term one has to treat carefully is the gradient energy
\begin{align}
    E_\text{grad}=\frac{1}{2}\sum_{\vect{x}} \sum_{i=1}^{3} \Delta^{+}_{i}\phi \Delta^{+}_{i}\phi\,,
\end{align}
where $\Delta^{\pm}_i\phi$ denote the components of the forward and backward gradient
\begin{align}
    \Delta^{\pm}_{i}\phi(\vect{x})=\pm\frac{\phi(\vect{x}\pm dx\ \vect{\hat e}_i)-\phi(\vect{x})}{dx}\,,
\end{align}
where $dx$ is the spacial lattice spacing and $\vect{\hat e}_i$ the unit vector in direction $i$.
These reproduce $\nabla_i \phi(\vect{x})$ only up to $\mathcal{O}(dx)$ but $\nabla_i \phi(\vect{x}\pm dx/2\  \vect{\hat e}_i)$ to $\mathcal{O}(dx^2)$. We therefore employ the following averaging scheme to get an energy density that is correct up to $\mathcal{O}(dx^2)$ 
\begin{align}
    \rho_\text{tot}(\vect{x})=\frac{1}{2}\sum_{\Phi\in\{\phi,\psi\}}\left[\dot\Phi\dot\Phi + \frac{1}{2}\sum_{i=1}^{3}(\Delta^{+}_{i}\Phi \Delta^{+}_{i}\Phi+\Delta^{-}_{i}\Phi \Delta^{-}_{i}\Phi) \right] + V(\phi,\psi)\,.
\end{align}
This scheme has the added benefit of reproducing the energy that is used in \verb|CosmoLattice| when consistently evolving the scale factor or checking energy conservation
\begin{align}
    E_\text{tot}=\sum_{\vect{x}} \rho_\text{tot}(\vect{x})\,.
\end{align}
The expressions given here and below only hold in a flat space-time but can easily be generalized to expanding backgrounds using the $\alpha$-time concept of \verb|CosmoLattice| \cite{Figueroa:2020rrl,Figueroa:2021yhd}.

\subsection*{Shear}
The shear is the longitudinal-traceless component of the anisotropic stress. For a scalar field it is given as the outer product of the gradient \cref{eq:shear_scalarField} and we have to find an averaging scheme again to achieve $\mathcal{O}(dx^2)$ accuracy, since $\nabla_i\phi$ and $\nabla_j\phi$ are positioned on different sites of the lattice for $i\neq j$. In principle a scheme where one averages after taking the product is possible but we choose to go with a scheme using 
\begin{align}
    \Delta^\text{sym}_{i}\phi(\vect{x})=\frac{1}{2}\frac{\phi(\vect{x} + dx\ \vect{\hat e}_i)-\phi(\vect{x} - dx\ \vect{\hat e}_i)}{dx}\,,
\end{align}
since this is used for the simulation of gravitational waves in \verb|CosmoLattice| and the involved transverse-traceless projections \cite{Figueroa:2011ye}. We then have
\begin{align}
    T_{ij}\approx\sum_{\Phi\in\{\phi,\psi\}}\Delta^\text{sym}_{i}\Phi\Delta^\text{sym}_{j}\Phi\,.
\end{align}
After going to Fourier space $T_{ij}(\vect{k})$ with $\vect{k}=dk\ \vect{n}$ with $dk$ the infrared cut-off of the lattice and $\vect{n}\in \mathds{Z}^3$, we apply the following projector to find the shear
\begin{align}
     \tilde{\sigma}(\vect{k})=\frac{1}{\overline\rho}\sum_{i,j=0}^{3}\left(\vect{\hat p}^\text{sym}_i(\vect{k})\vect{\hat p}^\text{sym}_j(\vect{k})-\frac{1}{3}\delta_{ij}\right)T_{ij}(\vect{k})\,,
\end{align}
where multiplying by $-i\vect{p}^\text{sym}_i(\vect{k})$ in Fourier space corresponds to applying $\Delta^\text{sym}_{i}$ in position space
\begin{align}
    \vect{p}^\text{sym}_i(\vect{k})=\frac{1}{dx}\sin\left(dx\,k_i\right)\,.
\end{align}

\bibliographystyle{JHEP}
\bibliography{references}
\end{document}